\documentclass[twocolumn]{aastex6}
\usepackage{graphicx,psfig,amssymb,float,natbib,txfonts}

\begin{document}

\title{A {\em Chandra} X-ray census of the interacting binaries in old
  open clusters -- Collinder 261 }

\shorttitle{A {\em Chandra} study of the old open cluster Collinder 261}
\shortauthors{Vats \& van den Berg}

\author{Smriti Vats\altaffilmark{1} and Maureen van den Berg\altaffilmark{2,1}}
\affil{\altaffilmark{1}Anton Pannekoek Institute for Astronomy, University of Amsterdam, 
Science Park 904, 1098 XH Amsterdam, The Netherlands; S.Vats@uva.nl\vspace{0.5cm}}
\affil{\altaffilmark{2}Harvard-Smithsonian Center for Astrophysics, 60
  Garden Street, Cambridge, MA 02138, USA; mvandenberg@cfa.harvard.edu}

\begin{abstract}
We present the first X-ray study of Collinder 261 (Cr\,261), which at
an age of 7 Gyr is one of the oldest open clusters known in the
Galaxy. Our observation with the {\em Chandra X-ray Observatory} is
aimed at uncovering the close interacting binaries in Cr\,261, and
reaches a limiting X-ray luminosity of $L_X \approx 4 \times 10^{29}$
erg s$^{-1}$ (0.3--7~keV) for stars in the cluster. We detect 107
sources within the cluster half-mass radius $r_h$, and we estimate
that among the sources with $L_X \gtrsim 10^{30}$ erg s$^{-1}$,
$\sim$26 are associated with the cluster. We identify a mix of active
binaries and candidate active binaries, candidate cataclysmic
variables, and stars that have ``straggled'' from the main locus of
Cr\,261 in the colour-magnitude diagram. Based on a deep optical
source catalogue of the field, we estimate that Cr\,261 has an
approximate mass of 6500 $M_{\odot}$, roughly the same as the old open
cluster NGC\,6791. The X-ray emissivity of Cr\,261 is similar to that
of other old open clusters, supporting the trend that they are more
luminous in X-rays per unit mass than old populations of higher
(globular clusters) and lower (the local neighbourhood) stellar
density. This implies that the dynamical destruction of binaries in
the densest environments is not solely responsible for the observed
differences in X-ray emissivity.
\end{abstract}


\section{Introduction} \label{sec_intro}

Open clusters with ages in excess of a few Gyr are relatively rare in
the Galaxy (e.g.~\citealt{Kharchenko:2013p776}). Some aspect of their
properties (perhaps their large initial mass or their location out of
the Galactic plane, where they avoid interactions with large molecular
clouds or the disruptive pull of external gravitational forces) helped
them survive until old age. Studies of old open clusters, with their
well-developed sub-giant and giant branches, have been a cornerstone
of stellar-evolution theory for many decades, thanks, in part, to
their accurately measured ages and distances.

From the X-ray point of view, old open clusters are interesting for a
number of reasons. First, X-ray observations efficiently detect
different classes of close, interacting binaries, enabling the study
of processes such as tidal coupling and the link between X-rays and
rotation. The X-ray luminosity of late-type stars strongly depends on
stellar rotation. As single stars age, they spin down due to magnetic
braking \citep{Pallavicini:1989p1112}. As a result, their X-ray
emission decreases accordingly. An old star like our Sun ($\sim$4.5
Gyr) has an X-ray luminosity of about 10$^{26}$ to 10$^{27}$ erg
s$^{-1}$ (0.1--2.4~keV; \citealt{Peres:2000p1107}). Even with the
deepest exposures of a sensitive X-ray telescope like the {\em Chandra
  X-ray Observatory}, this is nearly impossible to detect except for
the nearest stars. Nevertheless, an early {\em ROSAT} observation of
the old open cluster M\,67, which lies at $\sim$840 pc
\citep{Pasquini:2008p1081} and is about as old as the Sun ($4\pm0.5$
Gyr; \citealt{Dinescu:1995p1080}), revealed a large number of X-ray
sources among the cluster members \citep{Belloni:1993p1075}. Many of
these turned out to be close, tidally interacting binaries where the
stellar rotation is locked to the orbital period, and therefore kept
at a level that can sustain magnetically active coronae. Subsequent
{\em XMM-Newton} \citep{Gondoin:2005p1051, Giardino:2008p1048,
  Gosnell:2012p685}, and $Chandra$ \citep{vandenBerg:2004p1040,
  vandenBerg:2013p442, Giardino:2008p1048} observations of old open
clusters have detected many such active binaries (ABs). ABs can be
binaries of two detached stars, or they can have a contact or
semi-detached configuration such as in W\,UMa and Algol binaries,
respectively. In terms of number of sources, ABs are the most
prominent X-ray source class in old open clusters, but other classes
of interacting binary are represented as well. In cataclysmic
variables (CVs), the X-rays are the result of accretion from a
late-type main-sequence donor onto a white dwarf. In fact, the first
{\em ROSAT} observation of M\,67 was aimed at studying the X-rays from
a CV that was discovered in the optical
\citep{Gilliland:1991p1033}. The origin of the X-ray emission from
more exotic open-cluster binaries, like blue stragglers, is less well
understood, but in X-rays they are more similar to the ABs than to the
mass-transfer sources \citep{vandenBerg:2013p1013}.

A second motive for studying old open clusters in X-rays, is that
their stellar densities lie in between those of the solar
neighbourhood ($\sim0.1 M_{\odot}$ pc$^{-3}$) and dense globular
clusters ($\geq 10^4 M_{\odot}$ pc$^{-3}$). This allows an
investigation of the effect of stellar dynamics on the clusters'
close-binary population, in a poorly studied density regime. With the
growing sample of old open clusters studied in X-rays, it is now
possible to do simple statistics regarding the number of sources
detected in each source class. It was found that the number of CVs in
M\,67 and NGC\,6791 scale with the present-day cluster mass, pointing
at a primordial origin. For ABs, that proportionality is not so
obvious, raising the issue of whether dynamical interactions that
break up or create binaries, play a role
\citep{vandenBerg:2013p442}. The expected low encounter rates in open
clusters do not seem to favour the latter explanation. Nevertheless,
there are clues that dynamical encounters shape the properties of at
least some binaries. N-body models of M\,67 \citep{Hurley:2005p1005}
suggest that primordial binaries {\em and} dynamical encounters are
necessary to explain the blue-straggler population of M\,67.
Some individual systems, such as the likely triple S\,1082 in M\,67
\citep{vandenBerg:2001p1001, Sandquist:2003p999} is also difficult to
explain without invoking encounters. Therefore, the origin of the
X-ray sources of old open clusters may not be solely primordial.

The X-ray emissivity, or the integrated X-ray luminosity per unit of
mass, of globular clusters is lower than that of M\,67 after removing
the contribution from luminous low-mass X-ray binaries (LMXBs;
e.g.~\citealt{verb01}). \citet{geea15} compared the X-ray emissivities
of more diverse environments including dwarf elliptical galaxies and
the local neighbourhood, and found that old open clusters also have
higher X-ray emissivities than other old stellar populations. Various
explanations have been suggested, relating to either the overall
mass-loss history of the clusters, differences in dynamical encounter
rates, or the processes underlying the X-ray emission. More study is
needed to determine which of these factors are responsible.

In order to improve the census of X-ray sources in old open clusters,
we are undertaking a survey with {\em Chandra} of open clusters with
ages between 3.5 and 10 Gyr. The observations are designed to reach a
limiting luminosity of $L_X\approx10^{30}$ erg s$^{-1}$ (0.3--7~keV),
or better, at the distance of the clusters. As part of this survey, we
have carried out the first X-ray study of Collinder 261 (Cr\,261), and
we present the results of our efforts in this paper. With an
estimated age of 6--7 Gyr \citep{Bragaglia2006p131}, Cr\,261 is one of
the oldest open clusters in the Galaxy, being superseded in age by
NGC\,6791 (8--9 Gyr) and Berkeley\,17 (8.5--10 Gyr) only. The cluster
metallicity is close to solar (\citealt{drazea16}), and reported
values for the distance and reddening lie between 2.2--2.7 kpc and
$E(B-V) \approx 0.25-0.34$, respectively (see
e.g.~\citealt{Gozzoli:1996p186}, \citealt{Carraro:1999p997},
\citealt{Bragaglia2006p131}), with a higher value of the reddening
considered more plausible \citep{frieea03}. In this paper, we adopt a
distance of 2.5 kpc and $E(B-V)=0.34$, unless stated otherwise. The
latter corresponds to a $V$-band extinction $A_V=1.05$ for the
canonical ratio $A_V/E(B-V)=3.1$, and a neutral hydrogen column
density $N_H=1.9 \times 10^{21}$ cm$^{-2}$ \citep{Predehl:1995p1114}.
The Galactic coordinates of Cr\,261 are $l=301.7^{\circ}$,
$b=-5.5^{\circ}$; due to its low Galactic latitude and location
towards the bulge, the number of fore- and background stars projected
onto the cluster is high. Cluster membership is poorly constrained for
the majority of stars in the field. Cr\,261 is included in the star
cluster catalogue of \citet{Kharchenko:2013p776}, which lists
structural parameters such as the overall size of the cluster and the
radius of its central region. In this work, we present an estimate for
the half-mass radius $r_h$ and the approximate mass of Cr\,261, which,
to our knowledge, have not been reported in the literature
before. These parameters facilitate a uniform comparison with the
X-ray properties of other old Galactic clusters.

We present the X-ray and optical observations, and the data reduction
in Sect.~\ref{sec_obs}. In Sect.~\ref{sec_ana} we describe the
analysis, which includes the creation of the X-ray and optical source
catalogues, their cross-correlation to identify candidate optical
counterparts to the {\em Chandra} sources, and the derivation of the
structural properties of Cr\,261. Sect.~\ref{sec_res} is focused on
the X-ray source classification. In Sect.~\ref{sec_disc} we discuss
our results in the context of the X-ray emission from other old
stellar populations, and we summarise our findings in
Sect.~\ref{sec_sum}.

\begin{figure}
\includegraphics[clip=,width=1.0\columnwidth]{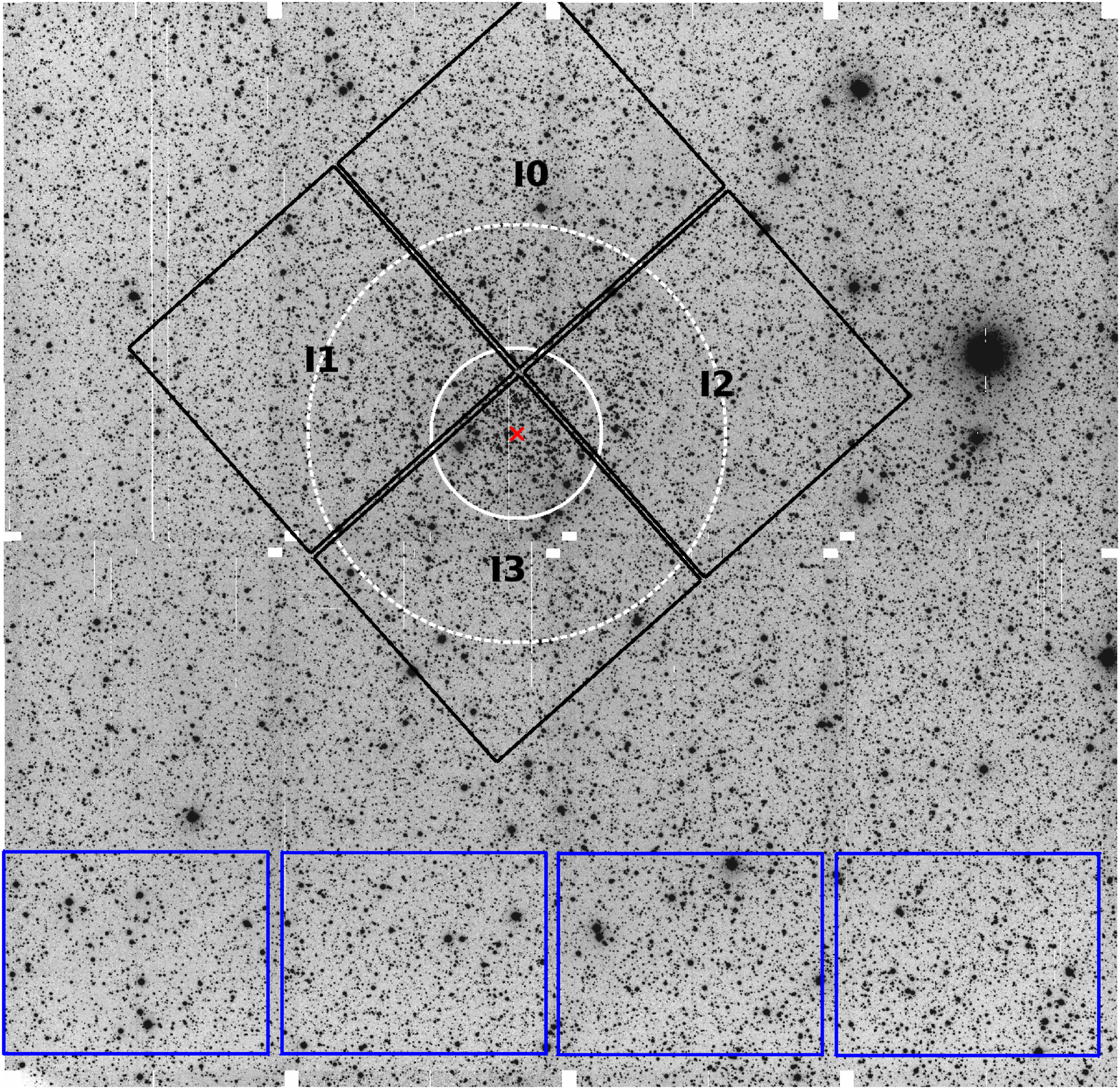}
\caption{Stacked WFI $V$-band image of Cr\,261. The four black squares
  show the ACIS-I field of view with chip IDs marked. The solid white
  circle marks the core radius of the cluster
  ($r_c=157\arcsec\pm16\arcsec$), centred on the cluster centre
  (marked by the red cross) as determined by us (see
  Sect.~\ref{sec_clusterprop}). The dashed white circle marks the
  half-mass radius of the cluster ($r_h=384\arcsec\pm38\arcsec$). Blue
  rectangles show the offset field used for determining the background
  stellar density. Small white rectangles are regions of zero optical
  exposure. North is up, east to the left.}
\end{figure}
\label{fov}

\section{OBSERVATIONS AND DATA REDUCTION} \label{sec_obs}

\subsection{X-ray Observations}  

Cr\,261 was observed with the Advanced CCD Imaging Spectrometer
\citep[ACIS;][]{Garmire:2003p984} on board \textit{Chandra} starting
2009 November 9 14:50 UTC for a total exposure time of 53.8 ks (ObsID
11308). The observation was made in Very Faint, Timed exposure mode,
with a single frame exposure time of 3.2 s. \citet{Kharchenko:2013p776}
estimate that the radius\footnote{Here we refer to the
  \citet{Kharchenko:2013p776} parameter $r_2$, which is defined as the
  distance from the cluster centre where the projected stellar density
  drops to the average stellar density of the field.}  of Cr\,261 is
$\sim$14\farcm1. This is considerably larger than a single ACIS chip
(8$\farcm$4$\times$8$\farcm$4); therefore, we placed the centre of the
cluster ($\alpha_{2000} = 12^{\rm h}38^{\rm m}06\fs0$, $\delta_{2000}
= -68^{\circ}22\arcmin01\arcsec$;
\citealt{Kharchenko:2013p776}\footnote{The cluster centre is
  redetermined in Sect.~\ref{sec_clusterprop}.})  close to the I\,3
aimpoint, so that a larger contiguous part of the cluster could be
imaged (see Figure \ref{fov}). The CCDs used were I\,0, I\,1, I\,2 and
I\,3 from the ACIS-I array, and S\,2 and S\,3 from the ACIS-S array.

We started the data reduction with the level-1 event file produced by
the data processing pipeline of the \textit{Chandra} X-ray Center and
used CIAO 4.5 with CALDB 4.5.5.1 calibration files for further
processing. To create the level-2 event file we used the
\verb|chandra_repro| script. A background light curve in the energy
range 0.3--7~keV was created with the CIAO \verb|dmextract| routine
using source-free areas on the ACIS-I chips, and was analysed with the
\verb|lc_sigma_clip| routine. No background flares with more than
3$\sigma$ excursions from the average background count rate were
observed, hence the total exposure was used for further analysis.

\subsection{Optical Observations} 

We retrieved optical images of Cr\,261 in the $B$ and $V$ bands from
the ESO public archive. These data were taken as part of the ESO
Imaging Survey (EIS; program ID 164.O-0561). The observations of
Cr\,261 were made using the Wide Field Imager (WFI), mounted on the
2.2m MPG/ESO telescope at La Silla, Chile. The WFI has a field of view
of 34$\arcmin$$\times$33$\arcmin$ covered by a detector array of eight
2k$\times$4k CCDs with a pixel scale of 0\farcs238 pixel$^{-1}$. The
Cr\,261 data were taken from 2001 June 27 23:55 UTC to 2001 June 28
00:38 UTC, with a total exposure time of 510 s in the $B$ and $V$
filter each. In each filter, two exposures of 240 s were taken,
supplemented with a single short exposure of 30 s to get photometry
for the bright stars. We only used the long exposures for our
analysis. The seeing during the observations was $\sim$1$\farcs$15.

For reducing the optical images we used the Image Reduction and
Analysis Facility (IRAF\footnote{IRAF is distributed by the National 
Optical Astronomy Observatory, which is operated by the Association 
of Universities for Research in Astronomy (AURA) under a cooperative 
agreement with the National Science Foundation.}) v2.16, supplemented 
by the MSCRED package for handling and reducing mosaic data. Basic data 
reduction steps of bias subtraction and flat-fielding were performed using 
the bias, and dome-and sky-flat images taken within one day of the science
exposures. With the MSCCMATCH routine, the geometric distortion of the
images was removed, and the eight individual chips of a given exposure
were combined into a single image. We created a master $V$-band image
by stacking the two individual, slightly offset, 240-s $V$-band
exposures. As a result, in the stacked $V$ image the space between the
individual chips of the WFI mosaic (23\arcsec~wide along the length of
the chips, and 14\arcsec~wide along their width) is largely, but not
completely, filled in (see Figure \ref{fov}). The stellar profiles in
one of the $B$-band images of Cr\,261 are very distorted, which
prevented us from modelling a good PSF. Since this degrades the
quality of the derived photometry, we opted to discard this
low-quality image and use only a single 240-s $B$ exposure for our
analysis. Therefore, our $B$-band catalogue of the Cr\,261 field has
no coverage in the chip gaps.

\begin{figure*}
\centerline{
\\
\includegraphics[width=7cm,angle=90]{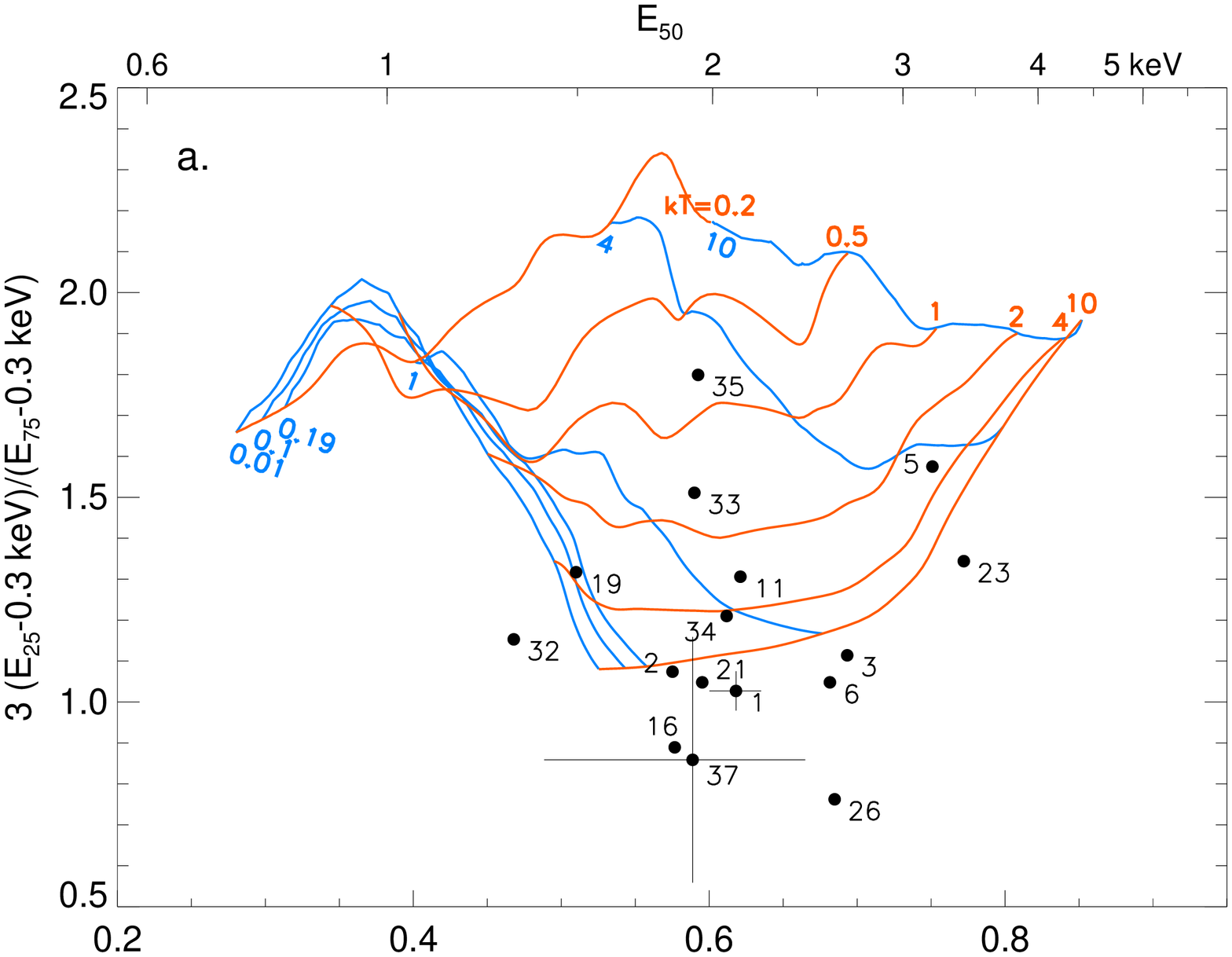}
\hspace{-0.5cm}
\includegraphics[width=7cm,angle=90]{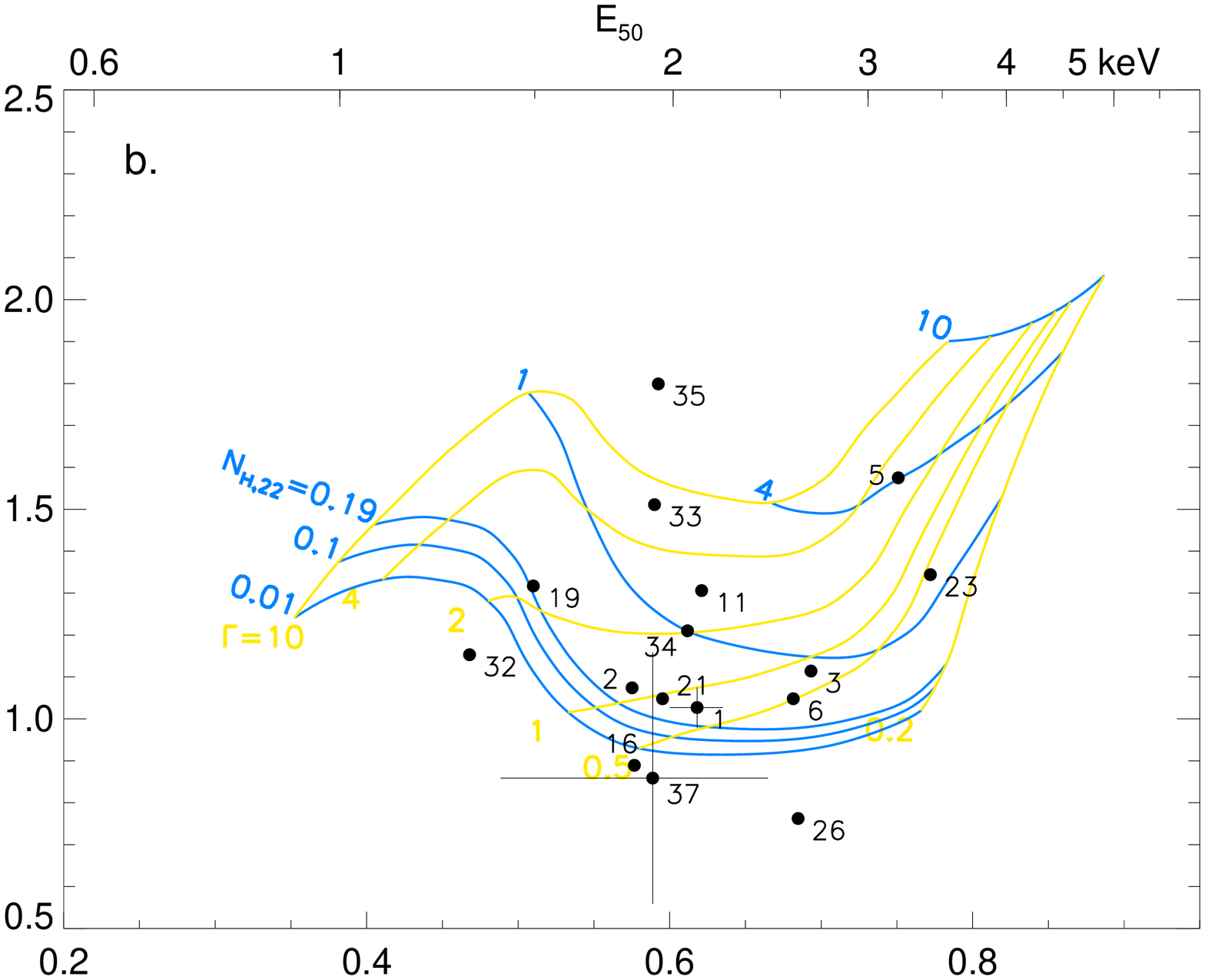}
}
%\end{figure*}
%\begin{figure*}
\centerline{
\\
\includegraphics[width=7cm,angle=90]{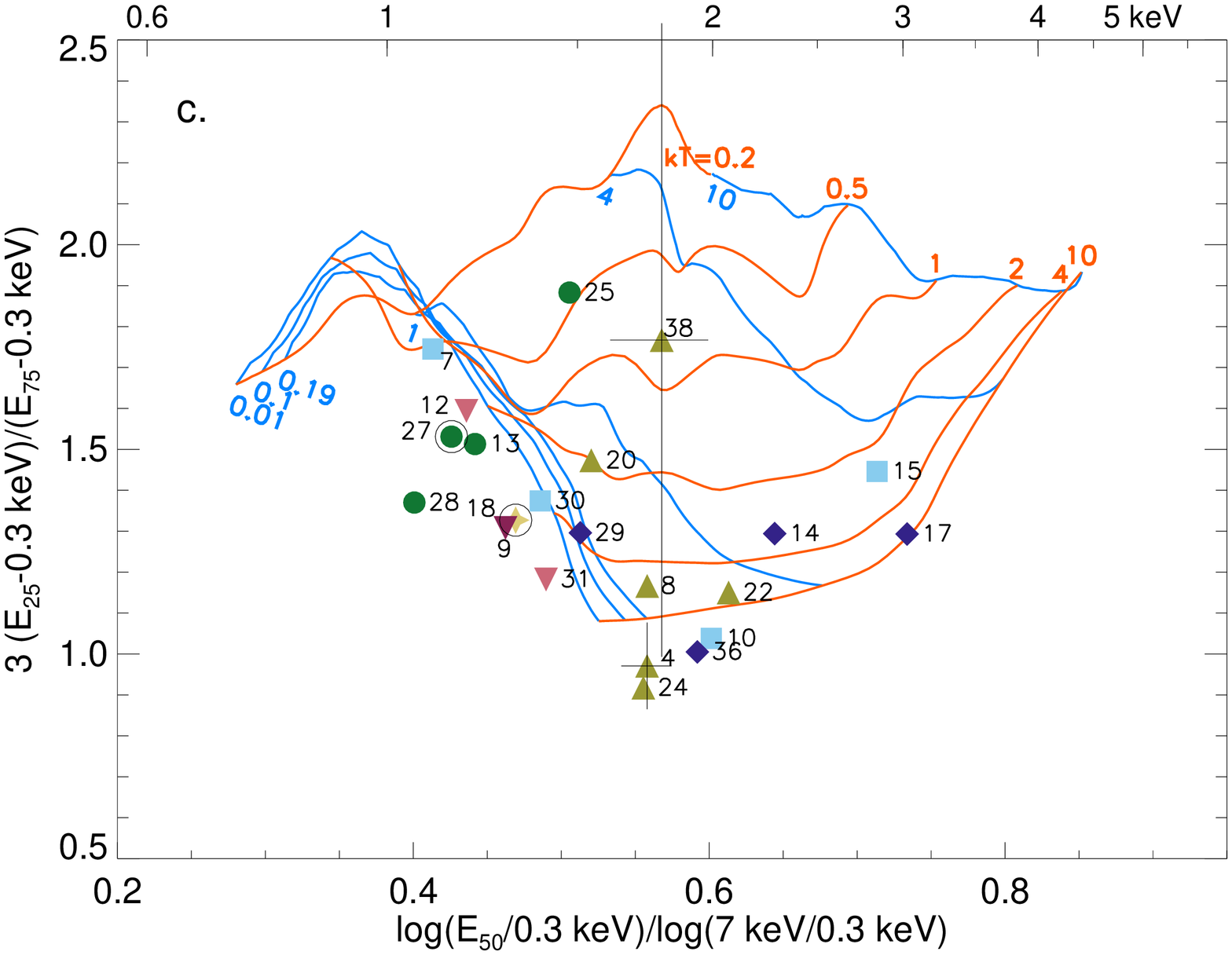}
\hspace{-0.5cm}
\includegraphics[width=7cm,angle=90]{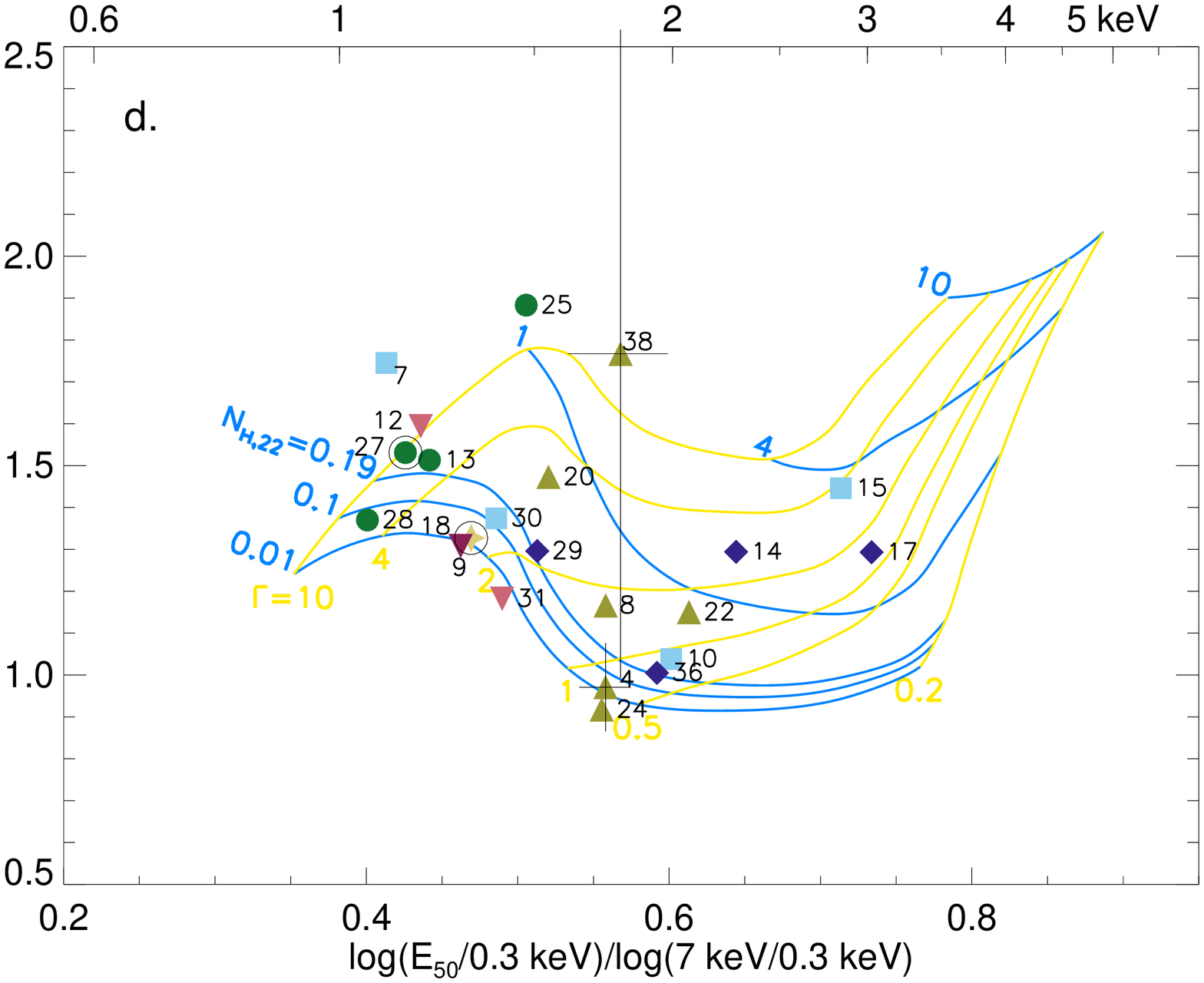}
}
\caption{Quantile diagrams with model grids representing
  a MeKaL plasma (left) and a power-law spectrum (right). The top
  panels show sources {\em without}, and the bottom panels show
  sources {\em with} candidate optical counterparts (see
  Table~\ref{tab2} for their classification). The plasma temperature
  $kT$ or photon index $\Gamma$, and the column density $N_H$ can be
  estimated from the location of a source with respect to the grid:
  blue curves represent lines of constant $N_{H}$ normalised in units
  of $10^{22}$ cm$^{-2}$ ($N_{H,22}$, where $N_{H,22}\approx 0.19$
  cm$^{-2}$ is the cluster value), whereas orange curves are lines of
  constant $kT$ (labeled in units of keV; left), and yellow curves
  are lines of constant $\Gamma$ (right). The median energy $E_{50}$
  can be read off from the top x-axis. Here we show sources with
  twenty net counts (0.3--7~keV) or more; error bars are shown only
  for the sources with the highest and lowest number of counts in a
  given panel. Filled coloured symbols mark X-ray sources for which
  we have found candidate optical counterparts. Among them, green circles represent ABs and
  candidate ABs (Sect.~\ref{sec_ab}), olive triangles are for
  candidate CVs or AGNs (Sect.~\ref{sec_cv}), yellow four-point stars
  for candidate BSSs, pale red downward triangles for candidate SSGs,
  maroon downward triangles for candidate YSS (Sect.~\ref{sec_bss}),
  pale blue squares for likely non-members of the cluster
  (Sect.~\ref{sec_nonmem}), and deep blue diamonds for sources with
  uncertain classification (Sect.~\ref{sec_unclass}). Furthermore, sources CX\,18 and CX\,27 that
  have close-binary counterparts \citep{Mazur:1995p95} are marked with
  a larger black open circle.}
 \label{quantiles}
\end{figure*}

\section{ANALYSIS} \label{sec_ana}

\subsection{X-ray Source Detection and Source Characterisation} \label{sec_xana}

We limited the X-ray analysis to the data from chips I\,0, I\,1, I\,2
and I\,3. The S\,2 and S\,3 chips lie far from the I3 aimpoint, giving
rise to large positional errors on any sources detected on them. Such
large errors make it hard to identify optical counterparts, and thus
to classify the sources.

Source detection was done in a soft (0.3--2~keV), hard (2--7~keV), and
broad (0.3--7~keV) energy band. The CIAO source detection routine
\verb|wavdetect| was run for eight wavelet scales ranging from 1.0 to
11.3 pixels, each increasing by a factor of $\sqrt{2}$. Larger scales
are better suited for more off-axis sources, where the PSF is wider or
more distorted. Exposure maps were computed for an energy value of
1.5~keV. The \verb|wavdetect| detection threshold ({\tt sigthresh} )
was set at $ 10^{-7} $. The corresponding expected number of spurious
detections per wavelet scale is 0.42 for all four ACIS chips combined,
or 3.35 in total for all wavelet scales. We ran \verb|wavdetect| for
the three different energy bands and then cross-correlated the
resulting source lists to obtain a master X-ray source list. We
detected 113 distinct X-ray sources. To check if we had missed any
real sources, we ran \verb|wavdetect| again for a detection threshold
of $ 10^{-6} $, which increases the expected total number of spurious
detections to 33.5. We found a total of 151 distinct X-ray sources
with more than two counts (0.3--7~keV) in this case. The positions of
seven of the extra 38 sources are found to match those of short-period
binaries discovered by \citet{Mazur:1995p95} (see
Sect.~\ref{sec_xomatch}). Close, interacting binaries are plausible
real X-ray sources and indeed the expected number of chance alignments
between the {\em Chandra} detections and the binaries in the Mazur
catalogue is very low (Sect.~\ref{sec_falsepos}). It is therefore
likely that at least these seven additional sources are real; but
given the $\sim$34 spurious detections that are expected, we do not
believe that there are many more real sources among the extra
detections. We flagged the sources that are only found for {\tt
  sigthresh}=$10^{-6} $, but kept them in the master source list.

For computing the positional uncertainties, required for
cross-correlation with other source catalogues, we used Eq.\,5 from
\citet{Hong:2005p959}, which gives the 95\% confidence radius on the
      {\tt wavdetect} position, $P_{err}$. The \verb|wavdetect|
      routine provides us with the source positions, but is not
      optimised to measure source counts. We determined the net source
      counts using ACIS Extract \citep[version
        2013mar6]{Broos:2010p966}. All events between 0.3 and 7~keV
      were extracted from regions enclosing $\sim$90\% of the PSF at
      1.5~keV. ACIS Extract also performs variability
      characterisation based on a Kolmogorov-Smirnov (K-S) test on the
      event arrival times for sources with five counts or more that
      spend more than 90\% of the total exposure time on the ACIS-I
      chips\footnote{based on the evaluation of the FRACEXPO keyword
        generated by {\tt mkarf} in CIAO}. For example, a source near
      a chip edge could effectively have a shorter exposure time if
      the telescope dither motion occasionally moves it off the
      detector. There were 76 sources with no evidence for
      variability $(0.05 < P_{KS})$; four sources which showed
      possible variability $(0.005 < P_{KS} < 0.05$; CX\,9, CX\,13,
      CX\,64, CX\,93) and four which were likely variable $(P_{KS} <
      0.005$; CX\,63, CX\,91, CX\,120, CX\,137), where $P_{KS}$ is the
      probability for a source to have a constant count rate. The
      X-ray light curves of the variables suggest flare-like
      behaviour, with a large fraction of the total events arriving in
      a relatively short time interval. The brightest of these sources
      is CX\,63, for which thirteen of seventeen events arrive in the
      last 3.5 h of the observation (and nine of seventeen events in a
      single hour). For the other three sources, 80\% or more of the
      events arrive within 2.5--3 h. X-ray flares are commonly
      observed in active late-type stars or binaries. This is
      consistent with our classification of CX\,120 (a W\,UMa binary
      and likely non-member of the cluster) and CX\,91 and CX\,137
      (likely foreground late-type dwarfs). The classification of
      CX\,63 is less secure, but it could be a late-type star or
      binary as well. These sources are further discussed in
      Sects.~\ref{sec_nonmem} and \ref{sec_unclass}.

Only five sources in our catalogue have more than 100 net counts
(0.3--7~keV), with the brightest source having 475 net counts. For the
majority of our sources the spectrum of the X-ray emission is
therefore poorly constrained. We calculated unabsorbed flux values in
the 0.3--7~keV band, $F_{X,u}$, for each source from its net count
rate and local {\tt rmf} and {\tt arf} response files using
\verb|Sherpa|. We assumed a 2~keV MeKaL model (\textit{xsmekal})
attenuated by a neutral hydrogen column density $N_{H}=1.9\times
10^{21}$ cm$^{-2}$ (the value for Cr\,261) using the {\em xstbabs}
model. The MeKaL model describes the emission from a hot, diffuse gas
or optically thin plasma, as is appropriate for ABs. Since the nature
of our sources is unknown a priori, and the number of counts is too
low to do any detailed spectral fitting, we explored the effect of
using different spectral models on the derived values of $F_{X,u}$. We
compared the unabsorbed flux values obtained using the 2~keV MeKaL
model with those obtained using a 1~keV MeKaL model, a 10~keV thermal
bremsstrahlung model (\textit{xsbrems}) and a power-law model
(\textit{xspowerlaw}) with a photon index, $\Gamma$, set to 1.4; the
\textit{xstbabs} model was used in all cases. The flux values obtained
using these models were about 6\% smaller, 40\% larger, and 80\%
larger, respectively, than the flux value obtained using the 2~keV
MeKaL model. The X-ray sensitivity limit, as defined by the unabsorbed
flux of the faintest detection, was found to be $\sim6\times10^{-16}$
erg cm$^{-2}$ s$^{-1}$ for the 2~keV MeKaL model and assumed cluster
$N_{H}$, which corresponds to an X-ray luminosity of
$L_X\approx4\times10^{29}$ erg s$^{-1}$ (0.3--7~keV) at the adopted
cluster distance (2.5 kpc).

In order to characterise the spectral properties of our X-ray sources
we use quantile analysis, which is optimised for sources with few
counts \citep{Hong:2004p940}. In this method, the median energy,
$E_{50}$, and 25\% and 75\% quartile energies ($E_{25}$ and $E_{75}$,
respectively) of the source events' energy distribution are used to
determine spectral hardness and spectral shape. Conventional X-ray
hardness ratios use fixed energy values for defining hard and soft
energy bands, and give results of little meaning if all events lie in
either the soft or the hard energy band. Details of the source
properties are presented in Table~\ref{tab1}, while quantile diagrams
are shown in Figure~\ref{quantiles} for sources with $\geq20$ net
counts (0.3--7~keV).

\begin{figure*}
\includegraphics[clip=,width=2.0\columnwidth]{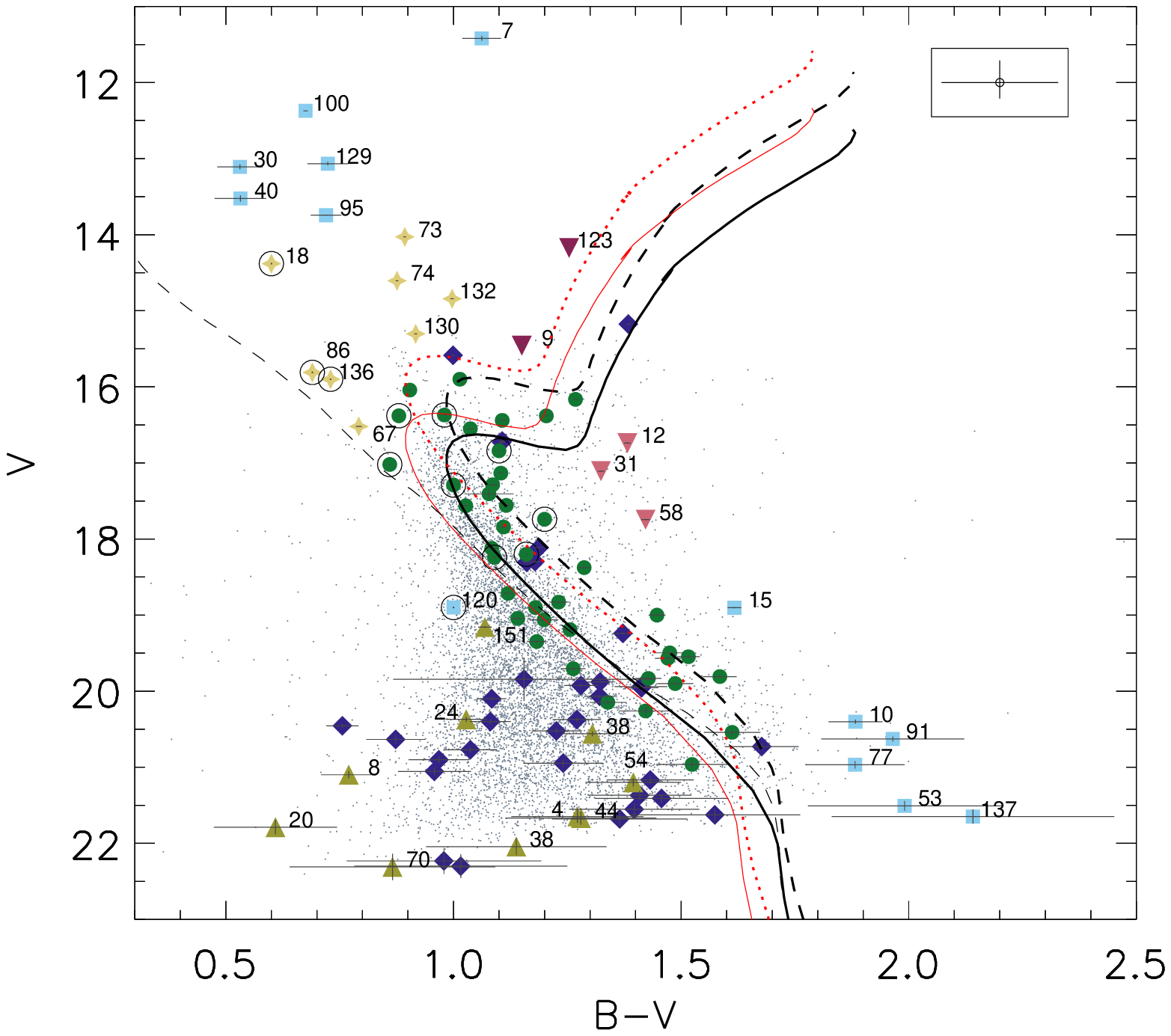}
\caption{Colour-magnitude diagram of Cr\,261 based on the WFI
  photometry. The different symbols and colours have the same meaning
  as in Figure \ref{quantiles}. Furthermore,
  sources with close-binary counterparts are circled with a larger
  black open circle; for these stars, the $BV$ photometry is obtained
  from \cite[see Sect.~\ref{sec_ab}]{Mazur:1995p95}. Solid lines
  represent isochrones
  \citep{Bressan:2012p1120,Chen:2014p1422,Tang:2014p1207,Chen:2015p452}
  for the upper and lower limits of the cluster reddening (red for
  $E(B-V)_{low}=0.25$, and black for $E(B-V)_{up}=0.34$). Dotted lines
  are the same isochrones but shifted upward by $-0.75$ mag to
  indicate the limit for unresolved photometric binaries. The dashed
  line represents a zero-age main-sequence isochrone. Error bars are
  smaller than the symbols for some sources. The combined uncertainty
  in the location of the isochrone due to reddening and distance
  uncertainties is shown in the rectangle at the top right. For
  clarity, the ABs and uncertain classifications have not been
  labeled.}
\label{hrd}
\end{figure*}

\subsection{Optical Source Catalogue} \label{sec_ocat}

The absolute astrometry of the optical images was tied to the
International Celestial Reference System (ICRS). We did this by
computing an astrometric solution based on the positions of
unsaturated stars in the field that are also included in the USNO CCD
Astrograph Catalog 4 (UCAC4; \citealt{Zacharias:2013p938}). For the
$V$ image we used 1912 unsaturated stars and obtained rms residuals
of 0$\farcs$141 in right ascension and 0$\farcs$166 in declination in
the solution. For the $B$ image, we used 1773 unsaturated stars and
obtained residuals of 0$\farcs$157 in right ascension and 0$\farcs$173
in declination.

For performing photometry, we used the \verb|DAOPHOT| package in
IRAF. After creating a source catalogue for the $B$ and the $V$ image
separately, we cross-matched each of them with the
\citet{Gozzoli:1996p186} catalogue in order to convert our
instrumental magnitudes to the Gozzoli et al.~calibrated magnitudes in
the Johnson system. The Gozzoli study covers a region of radius
3$\farcm$5 around their adopted cluster centre. We found 2018 matches
for the sources in the $B$ catalogue within a calibrated-magnitude
range 13.7 $< B <$ 24.0, and 2276 matches for the $V$ catalogue within
a range of 13.0 $< V <$ 22.5. We manually inspected all the matched
sources and found that none appeared to be blended or saturated. Over
these magnitude ranges, a constant offset provides a good
transformation from instrumental to calibrated magnitudes. The
resulting WFI source lists for the entire field have a calibrated
magnitude range of 12.9--23.5 in $V$ and 13.7--24.6 in $B$. Finally,
we cross-matched the $B$ and $V$ source lists to make a master optical
source list. Some sources in the master catalogue were detected in the
$V$ band but were not present in the single $B$ image due to its chip
gaps and a shorter exposure.

The colour-magnitude diagram (CMD) of Figure \ref{hrd} shows our $B$
and $V$ photometry of stars inside $r_h$ (see next section).

\subsection{Estimate for the Half-mass Radius and Mass of Cr\,261} \label{sec_clusterprop}

\begin{figure}
\includegraphics[clip=,width=1.0\columnwidth]{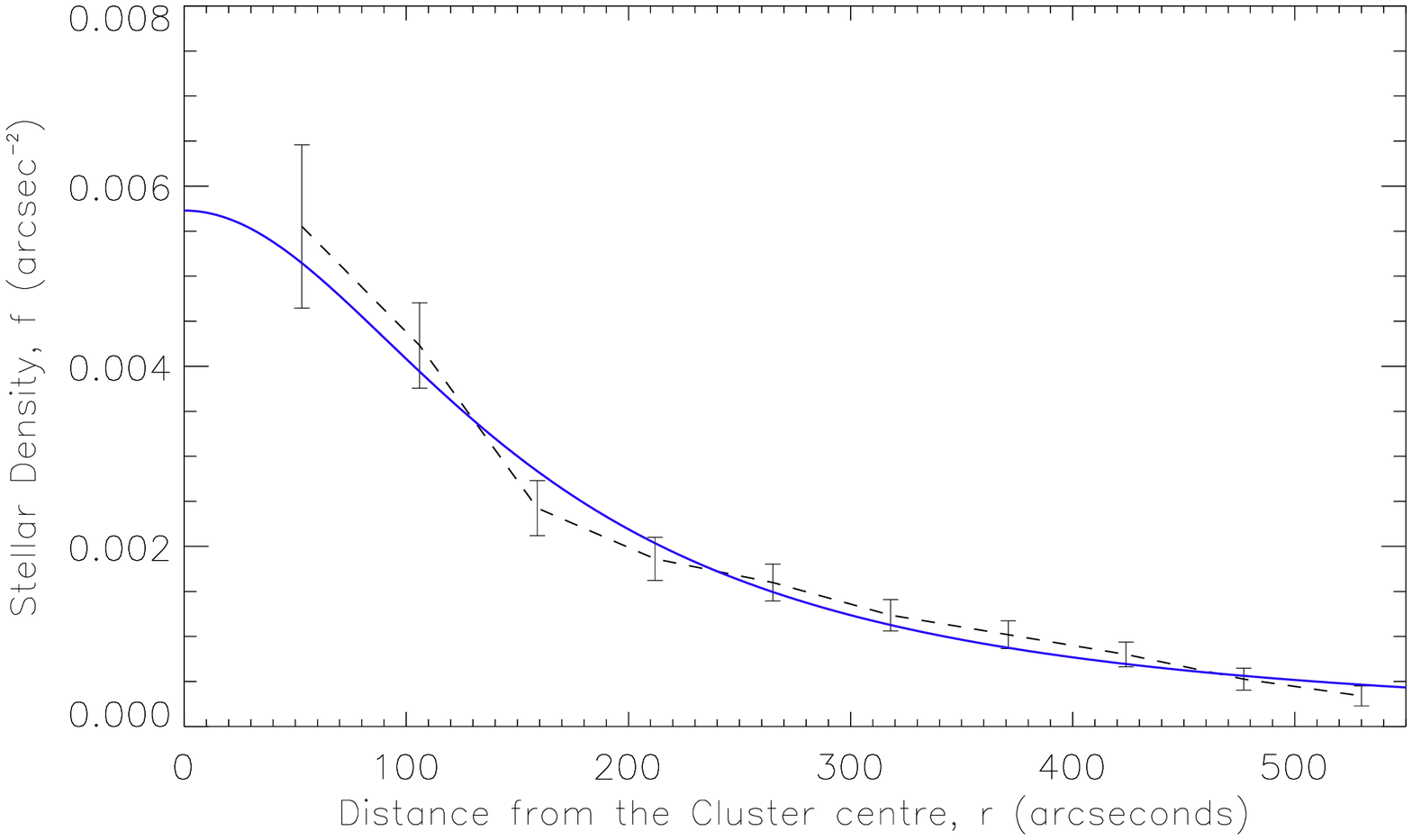}

\caption{Projected stellar density profile of Cr\,261 after correction
  for the contribution from foreground and background stars (points
  connected with a dashed line). The solid blue line shows the
  best-fitting King profile \citep{king62}, which has a central
  stellar density $f_0=0.0057\pm0.0007$ stars per arcsecond$^2$ and a
  core radius $r_c=157\arcsec\pm16\arcsec$.}
\label{stell_den}
\end{figure}

One of our aims is to compare the number of X-ray sources in Cr\,261
with those detected in other old open clusters and globular
clusters. Making a uniform comparison between clusters requires an
estimate for their masses and structural parameters. An estimate for
the King-profile \citep{king62} core radius $r_c$ of Cr\,261 was
derived by \citet{froeea10} and \citet{Kharchenko:2013p776}, who found
two significantly different values, viz.~52\arcsec~and
192\arcsec$\pm$48\arcsec, respectively. These values are based on the
2MASS near-infrared catalogue \citep{skruea06}, and (in case of
Kharchenko et al.~only) the optical PPMXL catalogue
\citep{roseea10}. Both star lists are relatively shallow, reaching
$\lesssim$1 mag below the Cr\,261 main-sequence turnoff. At the same
time, the PPMXL proper motions in the field of Cr\,261 (used by
Kharchenko et al.~to weed out possible non-members) have relatively
large errors ($\sim$9 mas yr$^{-1}$, on average) and do not display a
clear distinction between cluster stars and field stars. We decided to
derive our own estimate of $r_c$ and $r_h$, without making use of the
PPMXL proper motions.

In order to estimate $r_h$ for Cr\,261, we assumed that the stars are
symmetrically distributed about the cluster centre according to a King
profile. In Figure \ref{stell_den} we plot the projected stellar
density $f(r)$ versus radial offset from the cluster centre $r$,
computed in 50\arcsec-wide annular regions around the centre. Stars
were selected from the region between two 7-Gyr isochrones of solar
metallicity ($Z=0.019$; \citealt{Bressan:2012p1120}). One is modified
for a distance of 2.5 kpc and reddened by $E(B-V)=0.34$; the second
isochrone is the same, but shifted upward in the CMD by --0.75
mag. This is done to include the contribution from unresolved
photometric binaries (see Figure \ref{hrd}). To correct for the
contribution from stars that are unrelated to the cluster, we
estimated the density of stars within the same magnitude and colour
limits, from a catalogue we created from offset fields in our WFI
image (the blue rectangles in Figure \ref{fov}) that lie outside the
cluster radius $r_2$ \citep{Kharchenko:2013p776}. We fitted the
function $f(r)=f_0 (1+(r/r_c)^2)^{-1}$ to this background-corrected
radial profile; here $f_0$ is the central projected stellar
density. This function is the limit of the King profile for the
assumption that the tidal radius $r_t$ is much larger than $r_c$. In
the case of Cr\,261, the values of $r_c\approx0\farcm8$ and
$r_t\approx22\farcm2$ as derived by \citet{Kharchenko:2013p776},
support this assumption. Fitting the above function to the
radial-density profile, gives us a centre for the cluster that is
about 1\farcm5 different from the one given by Kharchenko et al.,
viz.~$\alpha_{2000} = 12^{\rm h}38^{\rm m}07\fs1$, $\delta_{2000} =
-68^{\circ}23\arcmin33\arcsec$, with a formal uncertainty of about
16\arcsec. We have used this new cluster centre for all purposes in
this study. The best-fitting King profile has
$r_c=157\arcsec\pm16\arcsec$, consistent with the value from
Kharchenko et al. Adopting this value of $r_c$ and assuming that the
total mass of the cluster is contained within $r_2$, we used Eq.~A3 in
\citet{Freire:2005p1474} to compute $r_h$. We find that
$r_h=384\arcsec\pm38\arcsec$.
 
We estimated the mass of Cr\,261 from the integrated $V$ magnitude
$I(V)$ of the cluster, following the method of
\citet{Bellazzini:2008p784}. We calculated the integrated magnitude of
stars inside $r_h$ (i.e.~$I_h(V)$) by summing the $V$-band fluxes of
the stars inside $r_h$ that satisfy the same magnitude and colour
restrictions as outlined above. Again, photometry of the offset fields
provides a correction for the flux density of foreground and
background stars. We converted $I_h(V)$ to the absolute integrated
$V$ magnitude of stars inside $r_h$, which resulted in
$M_h(V)\approx-3.6$. Next, we compared this value with the theoretical
curves for the evolution in time of the absolute $V$ magnitude of
solar-metallicity star clusters of various, constant, masses
\citep{Bellazzini:2008p784,Bragaglia:2012p778}. The age of Cr\,261 (7
Gyr) combined with our estimate for $M_h(V)$, yields an approximate
value for {\em half} the cluster mass of 4000--5500 $M_{\odot}$. The
uncertainty stems from the range spanned by the theoretical curves
computed for different initial-mass functions. As a final step, we
have reduced the inferred total mass (about 8000--11\,000 $M_{\odot}$)
with an empirical scaling factor. This was motivated by our finding
that the above method overestimates the masses of the old open
clusters M\,67 (by a factor 1.1--1.7) and NGC\,188 (by a factor of
1.3--1.9), for which accurate virial masses have been determined
\citep{gellea08,gellea15}. After scaling, our estimate for the {\em
  total} mass of Cr\,261 is about 5800--7200 $M_{\odot}$.

Obviously, our mass estimate should be considered as approximate,
only: we assumed that the total cluster mass is contained within
Kharchenko et al.'s cluster radius $r_2$, we have no comprehensive
list of members, and the evolutionary sequences for $M(V)$ from
\citet{Bellazzini:2008p784} may not be a perfect match to Cr\,261 (in
metallicity or mass function). For our purposes, though, this estimate
is good enough.

\subsection{Optical and X-ray Cross-matching} \label{sec_xomatch}

The possible error in the alignment of \textit{Chandra's} absolute
astrometry to the ICRS is small\footnote{For ACIS-I, the 95\%
  confidence radius on the alignment is
  $\sim$0\farcs9--1\arcsec~within a distance of 3\arcmin~from the
  aimpoint; see http://cxc.harvard.edu/cal/ASPECT/celmon}, but still
allows for a systematic offset between the X-ray positions and our
ICRS-calibrated optical positions. This systematic offset, or
boresight, can be comparable in size to the random errors on the X-ray
positions ($P_{err}$), and can complicate the search for optical
counterparts if not corrected for. To calculate the boresight, we used
the 45 short-period ($P < 3$ d) close binaries that were discovered by
\citet{Mazur:1995p95} in an optical-variability study of the Cr\,261
field. The reason for using the short-period variables for calculating
the boresight is that close binaries are plausible X-ray emitters and
hence, there is a lower chance of spurious detections that could
affect the boresight measurement (see also
Sect.~\ref{sec_falsepos}). With the finding charts in Mazur et al., we
were able to identify all 45 variables among the WFI sources. Their
WFI positions were then cross-matched with the X-ray catalogue, where
we adopted a 95\% match radius that combines the error in the optical
positions\footnote{\label{note5} The errors on the optical positions
  that are adopted here are the 1-$\sigma$ errors in the astrometric
  calibration given in Sect.~\ref{sec_ocat}, scaled to a 95\%
  confidence radius assuming a 2-D gaussian error distribution} and
the random error on the X-ray positions ($P_{err}$) in quadrature. To
account for errors in the alignment, we also add the 95\% confidence
radius on {\em Chandra's} absolute astrometry. Fifteen candidate
counterparts were thus found, which were then used to calculate the
boresight from the average X-ray--optical positional offsets. After
updating the X-ray positions for this initial boresight, the
cross-matching was repeated until the net boresight converged. This
method for calculating and correcting for the boresight is outlined in
detail in Sect.~3.3.1 of \citet{vandenBerg:2013p442}. We found a small
boresight that is consistent with zero, viz.~0\farcs06$\pm$0\farcs07
in right ascension and 0\farcs09$\pm$0\farcs08 in declination.

After correcting the X-ray source positions for the (almost
negligible) boresight, we matched our X-ray source list with the
entire optical source list, again using 95\% match radii. For 89
unique X-ray sources, we found 124 optical matches; of the latter, 104
are present in both the $V$ and $B$ images while for 20 we only have a
$V$ or $B$ detection. We also inspected the area around each X-ray
source in the WFI images by eye, and discovered that five more X-ray
sources have candidate optical counterparts that are saturated and
therefore missing from our optical catalogue. Finally, we added to the
list of candidate counterparts six optical sources that lie just
outside the 95\% match radius, but inside the 3-$\sigma$ radius. In
total, 98 of the 151 unique X-ray sources were thus matched to one or
more optical sources. For a complete list of candidate counterparts
and their optical properties we refer to Table~\ref{tab2}.

\subsection{False Positives Test, Background Galaxies, and Galactic Sources} \label{sec_falsepos}

To estimate the number of spurious matches between our X-ray and
optical sources, we calculated the surface density of optical
sources. Within $r_c$, the average density is 0.029 sources
arcsec$^{-2}$, while between $r_c$ and $r_h$ it drops slightly to
0.024 sources arcsec$^{-2}$. Multiplying the optical source densities
with the total area covered by the 95\% error circles of the X-ray
sources in the two regions, we expect 2.4 spurious matches among the
23 matches that we find in this central region, and 11.6 spurious
matches among the 47 matches in the outer region. Similarly, we use
the number of Mazur variables per arcsec$^2$ to estimate that the
number of spurious matches between X-ray sources and variables is
0.021 out of seven matches in the inner region, and 0.022 out of seven
matches for the outer region (one X-ray--detected Mazur variables lies
outside $r_h$). Therefore, all Mazur binaries that match with a {\em
  Chandra} source are likely real counterparts.

In order to estimate the number of background galaxies $N_B$ among our
X-ray detections, we used the relation for the cumulative number
density of high--galactic-latitude X-ray sources above a given flux
limit $S$ (Eq.~5 in \citealt{Kim:2007p933}). We adopted the $\log N_B
- \log S$ relation for the 0.3--8~keV band, which of the energy ranges
considered in the Kim study is closest to our broad band (0.3--7~keV). 
To convert counts to fluxes we adopted a power-law spectrum with
$\Gamma=1.4$ and $N_H=2.3 \times 10^{21}$ cm$^{-2}$, i.e.~equal to the
total integrated Galactic column density along the line of sight
\citep{marsea06}. We calculated $N_B$ for $r<r_c$, where most X-ray
sources that are truly associated with Cr\,261 are expected to
lie. The reason is that closer to the centre, the density of cluster
stars is simply higher; in addition, mass segregation makes the
radial distribution of binaries (and thus potential X-ray sources)
more concentrated. For a 5-count detection limit, we expect $N_B\approx9.5\pm3.1$ versus 22
sources actually detected. For a 10-count limit, it is expected that
$\sim5.7\pm2.4$ of the ten sources detected are extra-galactic. In the
region $r_c < r \leq r_h$, $45.7\pm6.8$ of the 58 sources detected above
5 counts, or $27.3\pm5.2$ of the 38 sources detected above 10 counts
are expected to be extra-galactic. These numbers indicate that we do
detect a population of, mainly faint, X-ray sources that is truly
associated with the cluster.

Given the low Galactic latitude of Cr\,261, a few foreground X-ray
sources are also expected to contaminate our sample. The exact number
is hard to estimate since there is no Galactic X-ray source density
distribution for this latitude that reaches down to our detection
limit. We have used the $\log N - \log S$ curves from Figure 9 in
\citet{Ebisawa:2005p1558} for the soft band (0.5--2~keV), and read off
the $\log N$ for a flux limit that corresponds to a 5-count detection
emitting a 2~keV MeKaL spectrum and $N_H=1.9\times10^{21}$
cm$^{-2}$. We expect $\sim$2.0 Galactic sources in the region inside
$r_c$ and $\sim$8.2 sources in the $r_c < r < r_h$ annulus; this must
be an upper limit since the Ebisawa field lies right in the plane
while Cr\,261 is a few degrees off. Other factors, such as the
difference between our and Ebisawa's soft band, and uncertainties in
the X-ray spectral model and $N_H$, also affect the accuracy of this
number.

\section{Results} \label{sec_res}

We used three criteria to classify our X-ray sources. First, we
considered the hardness of the X-ray spectrum as inferred from the
energy quantiles. Coronally active stars and binaries have thermal
X-ray spectra with plasma temperatures that generally do not exceed
$3-4$~keV \citep[e.g.][]{Gudel:2004p12}. The integrated Galactic
column density in the direction of Cr\,261 is $\sim$$2.3\times10^{21}$
cm$^{-2}$; Galactic X-ray sources without any intrinsic absorption
should therefore have an $N_H$ not larger than this. As a result of
these temperature and $N_H$ constraints, the expected $E_{50}$ values
for coronal sources are not much higher than $\sim$1.5~keV\footnote{In
  the MeKal grid of Figures \ref{quantiles}a and \ref{quantiles}c, the
  location $kT\approx4$~keV and $N_H\approx2 \times 10^{21}$ cm$^{-2}$
  corresponds to $E_{50} \approx 1.5$~keV (see top axes).}. On the
other hand, accreting binaries with compact objects, and AGNs often
have intrinsically harder X-ray spectra, and sometimes are observed
through additional, localised obscuring material; in both cases, the
expected $E_{50}$ is higher than $\sim$1.5~keV.

\begin{figure} 
\includegraphics[width=8cm]{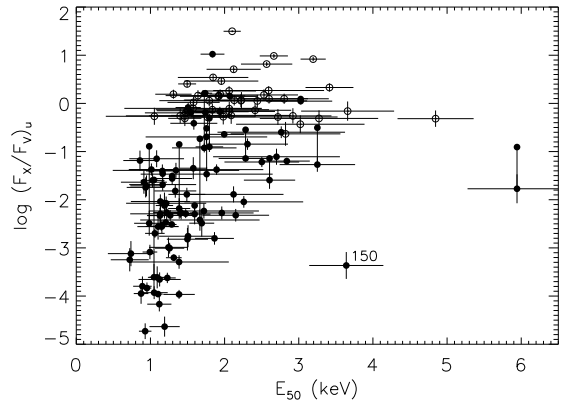}
\caption{The median energies $E_{50}$ and X-ray--to--optical flux
  ratios $\log (F_X/F_V)_u$ show a trend of lower flux ratios for soft
  sources than for harder sources. For sources without optical
  counterparts, the lower limit on $\log (F_X/F_V)_u$ (shown as open
  circles) was calculated for the detection limit $V=23.5$. For
  sources with multiple counterparts, the range on $\log (F_X/F_V)_u$
  is indicated with a vertical line. Error bars on the flux ratio only
  include statistical errors, not any systematic errors resulting from
  uncertainties in the adopted X-ray spectral model. CX\,150 is an
  outlier and may be spuriously matched to the star in its error
  circle. Only sources with $\sigma_{{\rm E}_{50}}$$<$ 1 are plotted.}
   \label{fig_e50_fxfo}
\end{figure}

Secondly, we looked at the ratio of the unabsorbed X-ray to optical
flux, or the limits thereon for sources without candidate optical
counterparts. We calculated this ratio with the equation $\log
(F_X/F_V)_u = \log F_{X,u} + V_0/2.5+5.44$, where the last term is the
logarithm of the $V$-band flux for sources with $V=0$. We adopted a
2~keV MeKaL model to calculate X-ray fluxes, assumed $N_H=1.9 \times
10^{21}$ cm$^{-2}$ to correct for absorption, and used
$V_0=V-A_V=V-1.05$. We caution that for most sources, $N_H$ is
unknown; if the adopted $N_H$ is lower (higher) than the actual $N_H$,
the flux ratio is overestimated (underestimated). Like $E_{50}$, the
flux ratio is mostly useful to distinguish between coronal and
accretion-powered sources. The former typically have $\log (F_X/F_V)_u
\lesssim -1$, with the most active late-type dwarfs reaching values of
about $-0.5$, while the latter have $\log (F_X/F_V)_u \gtrsim -1$
\citep{stocea91}. Indeed, for our sources, the average flux ratio is
lower for soft ($E_{50}\lesssim1.5$~keV) than for hard
($E_{50}\gtrsim1.5$~keV) sources (Figure \ref{fig_e50_fxfo}). An
optical source inside the X-ray error circle is not necessarily the
true counterpart, but can be a spurious match. Finding a relatively
hard X-ray source with a low $\log (F_X/F_V)_u$ value, can signal such
a random alignment.

For sources with candidate optical counterparts we also took into
account the position of these matches in the CMD. In most cases, this
works reasonably well to separate ABs from AGNs (which can lie far off
the cluster sequence) and CVs (which typically are blue). The position
in the CMD does a poor job in separating cluster stars from fore- or
background stars. As can be seen in Figure \ref{hrd}, and also in the
CMDs in \citet{Gozzoli:1996p186}, the cluster stars do not clearly
stand out. The lack of membership information for stars in the field
of Cr\,261 limits the classification of our {\em Chandra} sources, as
we discuss below. In the following, X-ray fluxes and luminosities
refer to the 0.3--7~keV band.

\subsection{Active Binaries and Candidate Active Binaries} \label{sec_ab}

For identifying possible ABs in Cr\,261, we selected sources with
candidate optical counterparts that lie along the cluster main
sequence or sub-giant branch; if a source has multiple matches that
all satisfy this condition, it is also classified as a (candidate)
AB. We allowed for the possible contribution to the light by a binary
companion, and for uncertainties in the reddening, as indicated by the
pairs of black and red isochrones in Figure \ref{hrd}. The uncertainty
in the cluster distance ($\sim$350 pc, based on the range of distances
reported in the literature), and in our absolute photometric
calibration (Sect.~\ref{sec_ocat}) are also sources of systematic
error (see the error bar in the top right of Figure
\ref{hrd}). Therefore, we classified candidate matches that are only a
little bit off the main sequence or sub-giant branch as ABs, too.
 
A total of 33 {\em Chandra} sources satisfy the photometric criteria
outlined above {\em and} have $E_{50}$ values within 1$\sigma$ of 1.5
~keV or lower; all have $\log (F_X/F_V)_u \lesssim -1.4$. We
classified them as ``AB'' in Table \ref{tab2}. Four additional
sources (CX\,62, CX\,80, CX\,115, CX\,125) have similar optical and
X-ray properties, but with errors on $E_{50}$ that are too large
($\gtrsim 1$~keV) to meaningfully constrain their X-ray spectra; these
sources were classified as ``AB?''. CX\,25 is an uncertain AB because
its position in the quantile diagram suggests an $N_H$ that is
enhanced with respect to the Galactic column, which is not expected
for typical ABs. Finally, CX\,41 and CX\,57 have $E_{50}=2.2\pm0.5$
~keV and $2.5\pm0.5$~keV, respectively; this is on the high side for
ABs, but given the large errors, we also put these two sources in the
``AB?''  category. With $B-V=1.34$ at $V=20.15$, CX\,41 is relatively
blue, but not as offset from the main sequence as the sources
discussed in Sect.~\ref{sec_cv}; however, it is not inconceivable that
this source is an AGN or a CV. We expect that a significant number of
``AB'' and ``AB?''  sources are fore- and background active stars or
binaries.

Ten ABs are matched to Mazur variables. CX\,27/V42, CX\,89/V11,
CX\,97/V38, and CX\,138/V21 are (semi-)detached eclipsing
binaries. The first three have periods of 1.3 d or shorter, while the
light curve of V21 shows eclipse-like events with an unconstrained
period. The maximum orbital period that can be tidally circularised in
$\sim$7 Gyr (i.e.\,the age Cr\,261) is $\sim$15 d
\citep{Mathieu:2004p602}. Since the time scale for tidal
synchronisation is shorter than that for circularisation
\citep{Hut:1981p99,Zahn:1989p220}, it is perfectly plausible that at
least V42, V11, and V38 contain rapidly rotating, and therefore
X-ray--active, stars. CX\,93/V30, CX\,126/V13,
CX\,133/V10\footnote{CX\,133 is matched with two stars on the main
  sequence; V10 is the more likely counterpart of the two.},
CX\,144/V24, CX\,147/V25, and CX\,149/V33 are contact binaries of the
W\,UMa type. For W\,UMa's, a distance constraint can be derived from
the known calibration of the absolute magnitudes in terms of orbital
period and $B-V$ or $V-I$ colour \citep[see
  e.g.][]{Rucinski:1997p109}. Mazur et al.~thus found that the
distances to V30, V13, V25, V33, and likely V24, are compatible with
that of Cr\,261; these are the most reliable cluster ABs in our
sample. Mazur et al.~were inconclusive regarding the distance to V10.

The time span of the WFI observations is $\sim$0.75 h, with the $B$
data taken first. For CX\,149/V33, with a period of 6.96 h and a
large-amplitude ($\sim$0.8 mag) light curve, this spans $\sim$0.1 in
orbital phase. If our observations happened to be timed around eclipse
ingress (something we cannot check because the ephemeris is not known
with sufficient precision), this can explain why we find a much bluer
colour ($B-V=0.72$) than Mazur et al., who report $B-V=1.09$,
i.e.~right on the main sequence. We found similar colour differences
for a few other variables. Mazur et al.~adopted a method that makes
their colours much less sensitive to non-simultaneous
measurements. Therefore, in Figure \ref{hrd}, we plotted the variables
with their Mazur photometry (see Table \ref{tab_mazur}), if
available.

\subsection{Candidate Cataclysmic Variables or AGN} \label{sec_cv}

Our mass estimate for Cr\,261 (5800--7200 $M_{\odot}$) is similar to
the mass of NGC\,6791 (5000--7000 $M_{\odot}$). If CVs in open
clusters are primordial, Cr\,261 would host a similar number of CVs as
NGC\,6791, i.e.~3 to 4 \citep{vandenBerg:2013p1013}. CVs typically lie
to the blue of the main sequence due to the light from the accretion
disk, and possibly from the white dwarf (although in the $B$ band, the
blue excess is not always that obvious, see
e.g.~\citealt{Bassa:2008p488}). For eleven sources, the candidate
optical counterpart(s) are blue with respect to the main sequence, and
ten of them are possibly CVs: CX\,4, CX\,8, CX\,20, CX\,22, CX\,24,
CX\,38, CX\,44, CX\,54, CX\,70, and CX\,151\footnote{The optical match
  to CX\,22, and one of the matches to CX\,24, are only detected in
  $B$, but the detection limit in $V$ implies they must be blue
  ($B-V\lesssim0.7$).}. The other blue source, CX\,120, is not a
member of the cluster (see Sect.~\ref{sec_nonmem}). We consider a
source to be blue if it lies to the left of the isochrone that is
reddened for the lowest possible cluster reddening. In addition, we
require the blueward offset from this isochrone to be at least 0.13
mag, i.e.~the errors on the absolute photometric calibration in $V$
and $B$ added in quadrature (see Sect.~\ref{sec_ocat}).

The ten sources listed above have $E_{50}$ values between $\sim$1.5
and 2.8~keV, $\log (F_X/F_V)_u$ between --2.5 and +1.0, and $L_X$
between $5\times10^{29}$ and $3\times10^{31}$ erg s$^{-1}$. This is
consistent with a CV classification, although $\log (F_X/F_V)_u
\approx -2.5$ for CX\,151 is on the low side for a CV; this may
suggest a different source class or the presence of a random
interloper in the X-ray error circle. We label these sources as
candidate CVs (``CV?''  in Table \,\ref{tab2}). Confusion with other
classes in this part of the CMD is mainly with AGNs, which outnumber
the CVs in the field observed (Sect.~\ref{sec_falsepos}) and can have
similar blue colours and X-ray properties. However, since very few
CVs in open clusters have been found, it is worthwhile to highlight
any candidates. Follow-up optical spectroscopy can confirm or disprove
whether a source is a CV nor not.

\subsection{Candidate Blue Stragglers, Yellow Stragglers, and Sub-subgiants} \label{sec_bss}

Some of the brightest X-ray sources in old open clusters are members
that lie off the main locus of the cluster in the CMD. These systems
challenge our understanding of binary evolution, and, in some cases,
we do not understand why they emit X-rays \citep{vandenberg:1999p347}.
Therefore, they deserve special attention.

Blue straggler stars (BSSs) are bluer and brighter than the
main-sequence turnoff (MSTO) of a coeval population. Their formation
scenarios must explain how these stars managed to continue core
hydrogen burning for a longer time than cluster stars of similar
mass. Mass transfer in a binary, direct collisions, and the merger of
the close inner binary in a hierarchical triple driven by the
Kozai-Lidov mechanism, are the three proposed formation channels
\citep{davi15}. For most BSSs, it is not clear which (if any) of these
channels applies. The detection of X-rays in a bona-fide cluster BSS
is a sign of ongoing binary interaction and thus provides a clue to
the current system configuration. There is no strict brightness limit
with respect to the MSTO that we can use to select candidate BSSs in
Cr\,261. In M\,67, which has one of the best-studied BSSs populations,
the brightest BSS \citep[F\,81; ][]{Leonard:1996p470} lies $\sim$2.7
mag above the MSTO in the $V$ band. We take the equivalent location in
the CMD of Cr\,261, i.e.~$V\approx14$, as a (somewhat arbitrary)
limit, and consider brighter stars to be non-members. We thus find
eight matches with candidate BSSs: CX\,18/V45, CX\,67, CX\,73, CX\,74,
CX\,86/V12, CX\,130, CX\,132, and CX\,136/V22. Except for CX\,67,
these sources are soft ($E_{50}\lesssim1.4$~keV) and all have $\log
(F_X/F_V)_u$ between --3.9 and --2.8. This is consistent with the
properties of ABs, and their X-rays are therefore likely the result of
magnetic activity. Indeed, three sources are matched to
(semi-)detached eclipsing binaries with periods between 1.1 and 2.1
d. V12 and V22 show Algol-type light curves. The idea of a possible
link between Algols and BSSs was already put forth by
\citet{McCrea:1964p128..147M}. In an Algol binary, the originally less
massive star is now observed to be the more massive one as a result of
the mass it received from its Roche-lobe filling companion---here we
may be seeing a BSS ``in the making''. It would therefore be
particularly interesting to determine if V12 and V22 are cluster
members.

The candidate counterparts of CX\,9 and CX\,123 lie between the BSSs
and red giants. Stars in this region of the CMD have been dubbed
yellow stragglers and may be BSS descendants. All yellow stragglers in
M\,67 are solid cluster members and X-ray sources
\citep{Belloni:1998p1070}. Their X-ray properties point at the
presence of magnetic activity, and the same appears to be the case for
CX\,9 and CX\,123.

Finally, sub-subgiants (SSGs) or red stragglers, lie below the
sub-giant branch or to the red of the base of the giant
branch. Whereas BSSs seem to have somehow managed to prolong their
main-sequence lifetime, SSGs resemble (sub-)giants that have evolved
from stars less massive than the turnoff mass. Little is known about
the evolutionary history that has led to their current CMD position
(see \citet{vandenBerg:2013p1013} for a summary). We see three
candidate SSGs in Cr\,261: CX\,12, CX\,31, and perhaps CX\,58. Their
X-ray properties are consistent with those of ABs. CX\,58 may be too
faint for a SSGs, but just as there is no ``bright'' limit for BSSs,
there is no well-defined ``faint'' limit for SSGs.

The alternative explanation for the sources discussed above is that
they are foreground stars. Assuming $N_H=0$, they could be early-G to
late-K foreground dwarfs at distances up to 1.6 kpc (Mamajek
2016\footnote{http://www.pas.rochester.edu/$\sim$emamajek/EEM\_dwarf\_UBVIJHK\_
  \mbox{colors}\_Teff.txt}) with $L_X \approx (4-9) \times10^{27}$ erg
s$^{-1}$.

\subsection{Cluster Non-members} \label{sec_nonmem}

CX\,120 and CX\,68 are matched to the variables V34 and V20,
respectively. V34 was classified as a W\,UMa binary behind the
cluster. The eclipsing binary V20 lies well to the red of the main
sequence, which makes it an unlikely cluster member. Their X-ray
properties are consistent with those of ABs. Six more X-ray sources
have very red counterparts: CX\,10, CX\,15, CX\,53, CX\,77, CX\,91,
and CX\,137. If we assume these are foreground ($N_H\approx0$)
late-type stars, their $B-V$ colours suggest they are mid- to
late-type M dwarfs at about 40--215 pc; this implies $L_X \approx
1\times10^{26}$ to $4\times10^{28}$ erg s$^{-1}$. CX\,91 and CX\,137
are variable in X-rays, which could point at flares, another signature
of coronal activity. The position in the quantile diagram of CX\,15
suggests an $N_H$ that is higher than the cluster value; it could be
an AGN. For CX\,10 and CX\,77, $E_{50}$ is relatively high and they as
well may be AGNs.

The soft ($E_{50}\lesssim1.4$~keV) sources CX\,7, CX\,30, CX\,40,
CX\,59, CX\,95, CX\,100 and CX\,129 have counterparts that are
brighter ($V=11.4-13.7$) than our adopted bright limit for blue
stragglers in the cluster. These are likely foreground stars. We also
consider CX\,55, which is matched to a star to the blue of the MSTO,
as a likely non-member. Their $B-V$ colours are consistent with those
of mid-F to mid-K dwarfs (for $N_H=0$). Using the corresponding
distance estimates (75--2400 pc), we find
$L_X\approx(0.1-20)\times10^{29}$ erg s$^{-1}$.

\subsection{Unclassified Sources} \label{sec_unclass}

Nineteen sources remain unclassified for various reasons. Six have
candidate counterparts that were only detected in $V$, so colour
information is lacking. Eleven sources have multiple counterparts with
very different optical properties, including some that were detected
in $V$ or $B$ only. CX\,14 is a moderately hard source matched to two
optical sources near the main sequence. Its $N_H$ appears to be higher
than the Galactic value (Figure \ref{quantiles}); it may be an AGN and
both optical matches could be spurious. Finally, CX\,150 is one of the
faintest detections; if the source is real, $E_{50}$ suggests it is a
very hard, or very absorbed, source (Figure \ref{fig_e50_fxfo}). The
match to the star on the sub-giant branch may be coincidental.

\subsection{Sources Without Candidate Optical Counterparts}

For 53 sources, we do not find any candidate optical
counterparts. With the detection limit of the WFI images
($V\approx23.5$), we can place lower limits on their
X-ray--to--optical flux ratios. These range from
$\log(F_X/F_V)_{u,lim} \approx -0.6$ for the faintest (CX\,146) to
$\log(F_X/F_V)_{u,lim} \approx 1.5$ for the brightest (CX\,1)
unmatched source. This is consistent with very active late-type dwarfs
and accretion-powered sources. The average $E_{50}$ for unmatched
sources is $2.1\pm0.5$~keV, versus $1.7\pm0.6$~keV for sources that do
have candidate counterparts (for detections with 10 or more
counts). Given that we expect many extra-galactic sources in our field
(see Sect.~\ref{sec_falsepos}), it is likely that most sources without
an optical match are AGNs.

\section{Discussion} \label{sec_disc}

\subsection{Comparison with Other Old Open Clusters}

In order to uniformly compare our results with the X-ray sources in
other old open clusters, we select those sources from the Cr\,261
X-ray catalogue that are inside $r_h$, and are brighter than $L_X
\approx 1 \times 10^{30}$ erg s$^{-1}$ (for cluster members). 
For a detection limit of $L_X\approx 1 \times 10^{30}$ erg s$^{-1}$ 
(0.3 --7~keV; 2~keV MeKaL model), 
about 57$\pm$8 of the 83 sources inside r$_{h}$ above this luminosity cutoff are 
extra-galactic. Consequently, we estimate that 26$\pm$8 sources in this area 
are associated with the cluster, and consider this an upper limit given 
the uncertain number of fore- and background Galactic sources.

Allowing for the limitations on our classification, we list in
Table~\ref{tab5} our best estimate for the number of candidate CVs,
SSGs, and ABs in Cr\,261 (based on the discussion in
Sect.~\ref{sec_res}), and three other old open clusters. The lower
limit on $N_{AB}$ in Cr\,261 is set by the two W\,UMa's inside $r_h$
that \citet{Mazur:1995p95} place at the distance of Cr\,261,
viz.~CX\,93/V30 and CX\,147/V25. Note that these classes do not
capture all source types observed, since some are not represented in
each cluster (such as BSSs). In the table we also list a revised mass
estimate for M\,67. Instead of $\sim$1100 $M_{\odot}$ \citep{richea98}
that we adopted in \citet{vandenBerg:2013p442}, we now use the virial
mass of $2100^{+610}_{-550}$ $M_{\odot}$ \citep{gellea15}. Within the
uncertainty, the total number of X-ray sources $N_X$ in Cr\,261 is
consistent with the number for NGC\,6791, a cluster of similar mass
and age. Considering all four clusters, $N_X$ scales with mass more
convincingly than before \citep{vandenBerg:2013p442}, now that we add
a fourth cluster (Cr\,261) and use the updated mass for M\,67---but
$N_{AB}$ in M\,67 remains on the high side. A scaling by mass is
expected if the X-ray sources predominantly trace a primordial
population of binaries. Since ABs are the largest constituent of the
X-ray sources, all or most of them are likely primordial, but whether
this is also true for all CVs and SSGs is difficult to say given the
small-number statistics.

By combining our mass estimate with the total X-ray luminosity of
cluster sources inside $r_h$, we compute the X-ray emissivity of
Cr\,261 (last column of Table~\ref{tab5}). We do not know which
sources inside $r_h$ are cluster members and which are not. Therefore,
we simply scale down the sum of the individual $L_X$ values by the
ratio of $N_X$ to the total number of sources detected, so a factor
$26/83 \approx 0.31$. Given the uncertainty in the membership, a more
sophisticated calculation is not really warranted. For NGC\,6791, the
emissivity was taken from \citet{vandenBerg:2013p442}, and for M\,67 we
updated the value for the new mass. We incorporated new counterpart
and membership information from \citet{platea13} in the numbers for
NGC\,6819; the range in emissivity reflects more and less conservative
assumptions on which counterparts are cluster members or not. The
conversion of the X-ray luminosities from the 0.2--10~keV band, as
given in \citet{Gosnell:2012p685}, to our adopted band of 0.3--7~keV
was done assuming a 2~keV MeKaL model. Uncertainties in the
emissivities are likely dominated by errors in the cluster masses (up
to $\sim$30\% for M\,67) and the unknown membership status of some
sources, especially if they are bright. Systematic uncertainties in
the X-ray luminosities themselves have less impact: in
Sect.~\ref{sec_xana} we estimated that the difference in flux for a
1~keV and 2~keV X-ray model is $\sim$6\%; given that most X-ray
sources are ABs, with coronal temperatures not too far from these
values, it is not likely that the choice of X-ray model affects the
total X-ray flux by more than $\sim$10\%. Allowing for these
uncertainties, the four open clusters listed in Table \ref{tab5} all
have similar X-ray emissivities, about $5\times 10^{28}$ erg s$^{-1}$
$M_{\odot}^{-1}$ (0.3--7~keV).

\subsection{Comparison with Other Old Stellar Populations}

It has already been pointed that extrapolating the scaling relation of
the number of X-ray sources by mass as seen in low-density globular
clusters, predicts no X-ray emitting close binaries in the even
lower-density open clusters \citep[see e.g.][]{Gosnell:2012p685} like
those listed in Table~\ref{tab5}. This is clearly in contrast with the
number of open-cluster X-ray sources actually observed. Our results on
Cr\,261 are in line with this trend; by adding another measurement to
the cluster sample, the perceived ``overabundance'' of X-ray sources
in open clusters is put on more solid ground.

The dearth of ABs and CVs is directly reflected in the lower X-ray
emissivity of both low-density and high-density globular clusters
compared to old open clusters
\citep{verb01,huanea10,vandenBerg:2013p442,geea15}. Even the emission
from quiescent LMXBs and millisecond pulsars, whose presence has so
far only been confirmed in globular clusters, cannot make up for
that. Some suggested explanations for the differences in X-ray 
emissivity relate to dynamical processes. A
higher fraction of the initial cluster mass may have been lost (in the
form of evaporating low-mass stars) from open clusters, as their
relaxation times are shorter than those of more massive globular
clusters. Also, the higher encounter rates in massive globular
clusters lead to the more efficient destruction of binaries, including
(relatively wide) RS\,CVn binaries that contribute a large fraction of
the X-rays from (some) open clusters. Other possible explanations
relate to the process that underlies the X-ray
emission. \citet{huanea10} remarked that open clusters are younger than
globulars, and that the faster-spinning young stars could be more
active in X-rays (here we note that the stellar rotation in binaries
is set by the orbital period, not the age). In addition, open clusters
have higher metallicities than globular clusters, and there are
indications that population-I ABs produce more X-rays than their
population-II counterparts \citep{ottmea97}.

Interestingly, in a broader comparison of X-ray emissivities of old
stellar populations, \citet{geea15} found that old open clusters not
only have higher X-ray emissivities than globular clusters, but also
than other old stellar populations without recent star formation, such
as dwarf ellipticals, the outer bulge of M\,31, and the solar
neighbourhood. In those environments, the stellar density is much
lower than in the cores of massive globular clusters, casting doubt on
whether differences in density are solely responsible for the difference in
X-ray output of old stellar populations. 

\section{Summary} \label{sec_sum}

With {\em Chandra} we have carried out the first X-ray study of
Cr\,261, one of the oldest open clusters known in the Galaxy. We
detected 151 X-ray sources down to a limiting luminosity of $L_X
\approx 4 \times 10^{29}$ erg s$^{-1}$ (0.3--7~keV) for stars in the
cluster. Analysis of deep optical $B$ and $V$ images yielded candidate
counterparts to 98 sources. Considering their X-ray and optical
properties, we were able to derive constraints on the nature of many
sources, despite the lack of membership information. Of the 107
sources inside $r_h$, five are CVs (or other compact binaries) or
AGNs. Another 34 sources are (candidate) ABs, and eleven match with
stars that possibly followed non-standard evolutionary paths in the
cluster environment (blue and yellow stragglers, sub-subgiants)---this
group is likely contaminated by fore- and background stars. The
remaining sources inside $r_h$ have no optical counterparts (39), have
ambiguous classifications (7), or match with stars that are very
bright or very red (11); we expect that most of these are not
associated with the cluster. Follow-up work on the Cr\,261 sources,
such as optical spectroscopy of the proposed counterparts, or
proper-motion studies, are now needed to further constrain the nature
and membership status of the X-ray sources in Cr\,261, and arrive at a
cleaner census of the close binaries.

We used our optical source catalogue to derive an approximate mass for
Cr\,261. The total number of X-ray sources inside $r_h$ above
$L_X=1\times10^{30}$ erg s$^{-1}$ (corrected for the extra-galactic
background contribution) compared to the number of X-ray sources in
other old open clusters, is roughly proportional with cluster
mass. This points at a dominant primordial origin of the
X-ray--emitting sources. Combining the mass with the total X-ray
luminosity of cluster sources, we have constrained the X-ray
emissivity of Cr\,261. The result, $\sim5\times10^{28}$ erg s$^{-1}$
$M_{\odot}^{-1}$ ($\sim$30\% uncertainty), agrees with that of the old
open clusters NGC\,6819, M\,67, and NGC\,6791. This supports earlier
findings that old open clusters are more luminous in X-rays than other
old stellar populations, such as the local neighbourhood and globular
clusters. Given that the frequency of dynamic encounters in globular
clusters and the field is widely different, one may expect that
dynamical destruction of binaries is not (solely) responsible for the
relatively suppressed X-rays from these environments. It is plausible
that the explanation for the high X-ray emissivity of old open
clusters must be sought in the open clusters themselves. Other old
open clusters included in our {\em Chandra} survey span a range of
ages (3.5--10 Gyr) and metallicities ($[$Fe/H$]$ between --0.5 and
+0.4); our future work will explore the impact of these parameters on
the X-ray emission of old stellar clusters.

\acknowledgments The authors would like to thank J.~Hong for help with
the computation of the energy quantiles, L.~Bedin for doing a
quick-look reduction of the optical WFI data, and R.~Wijnands for
comments on an early version of the manuscript. We are grateful to A. Bragaglia 
for sharing an optical catalog of Cr\,261 stars to aid in the photometric  calibration. Part of this work is
based on observations made with ESO Telescopes at the La Silla Paranal
Observatory under programme ID 164.O-0561. This work is supported by
{\em Chandra} grant GO0-11110X. S.V. acknowledges the support of NOVA 
(Nederlandse Onderzoekschool voor Astronomie)\\

{\it Facilities:} \facility{CXO} \facility{Max Planck:2.2m (WFI)}

%
% Table 1
%

\begin{table*}[h]
\caption{Catalogue of {\em Chandra} sources in Cr\,261}
\label{tab1}
\begin{center}
\leavevmode
\begin{tabular}{lcccccccclc} \hline \hline              
(1) & (2) & (3) & (4) & (5) & (6) &(7) & (8) & (9) &(10) &  (11) \\
CX & CXOU\,J & $\alpha$ (J2000.0) & $\delta$ (J2000.0) & Error & $\theta$ & C$_{t,net}$ & C$_{s,net}$ & $F_{X,u}$ & $E_{50}$ & Optical Match    \\ 
 & & (\arcdeg) & (\arcdeg) & (\arcsec) & (\arcmin) & & & (10$^{-15}$ erg cm$^{-2}$ s$^{-1}$) & (keV) & \\ \hline 
1 & 123823.4-682206 & 189.597301 & --68.368591 & 0.31 & 2.07 & 473$\pm$22 & 226$\pm$15 & 100 & 2.1$\pm$0.1 & -- \\ 
2 & 123740.9-682730 & 189.420561 & --68.458567 & 0.55 & 4.64 & 139$\pm$12 & 74$\pm$9 & 31 & 1.8$\pm$0.2 & -- \\ 
3 & 123854.1-681556 & 189.725375 & --68.265715 & 0.86 & 8.75 & 121$\pm$12 & 38$\pm$6 & 31 & 2.7$\pm$0.19 & -- \\ 
4 & 123902.9-682228 & 189.761982 & --68.374587 & 0.53 & 5.35 & 116$\pm$11 & 70$\pm$8 & 29 & 1.74$\pm$0.09 & + \\ 
5 & 123932.0-682559 & 189.883288 & --68.433159 & 1.06 & 8.18 & 101$\pm$10 & 17$\pm$4 & 27 & 3.19$\pm$0.17 & -- \\ 
6 & 123750.6-682807 & 189.460702 & --68.468808 & 0.63 & 4.83 & 94$\pm$10 & 32$\pm$6 & 21 & 2.6$\pm$0.3 & -- \\ 
7 & 123832.5-682949 & 189.635595 & --68.497007 & 0.92 & 6.70 & 93$\pm$10 & 90$\pm$10 & 24 & 1.10$\pm$0.06 & + \\ 
8 & 123839.7-681726 & 189.665606 & --68.290676 & 0.64 & 6.80 & 89$\pm$10 & 53$\pm$7 & 19 & 1.74$\pm$0.13 & + \\ 
9 & 123759.1-681609 & 189.496354 & --68.269261 & 0.76 & 7.42 & 71$\pm$9 & 52$\pm$7 & 16 & 1.29$\pm$0.07 & + \\ 
10 & 123744.9-682032 & 189.437242 & --68.342330 & 0.41 & 3.63 & 63$\pm$8 & 32$\pm$6 & 13 & 2.0$\pm$0.3 & + \\ 
11 & 123734.7-681418 & 189.394755 & --68.238507 & 1.35 & 9.70 & 61$\pm$8 & 29$\pm$6 & 17 & 2.1$\pm$0.4 & -- \\ 
12 & 123751.3-682535 & 189.463891 & --68.426524 & 0.46 & 2.51 & 61$\pm$8 & 54$\pm$7 & 14 & 1.18$\pm$0.07 & + \\ 
13 & 123757.7-682421 & 189.490520 & --68.406048 & 0.39 & 1.19 & 52$\pm$7 & 46$\pm$7 & 10 & 1.21$\pm$0.08 & + \\ 
14 & 123953.6-682051 & 189.973426 & --68.347770 & 2.03 & 10.18 & 50$\pm$8 & 23$\pm$5 & 12 & 2.3$\pm$0.4 & + \\ 
15 & 123931.8-681730 & 189.882528 & --68.291885 & 1.73 & 9.87 & 49$\pm$8 & 11$\pm$4 & 14 & 2.8$\pm$0.3 & + \\ 
16 & 123923.5-682200 & 189.848009 & --68.366711 & 1.06 & 7.21 & 45$\pm$7 & 24$\pm$5 & 11 & 1.8$\pm$0.5 & -- \\ 
17 & 123821.7-682820 & 189.590529 & --68.472426 & 0.84 & 4.99 & 44$\pm$7 & 13$\pm$4 & 11 & 3.0$\pm$0.4 & + \\ 
18 & 123816.6-682518 & 189.568966 & --68.421935 & 0.46 & 1.98 & 41$\pm$6 & 32$\pm$6 & 9.1 & 1.32$\pm$0.11 & + \\ 
19 & 123817.2-682039 & 189.571664 & --68.344429 & 0.40 & 3.02 & 41$\pm$6 & 30$\pm$5 & 8.2 & 1.50$\pm$0.12 & -- \\ 
20 & 123821.3-681312 & 189.588903 & --68.220081 & 1.99 & 10.42 & 40$\pm$7 & 27$\pm$6 & 10 & 1.55$\pm$0.15 & + \\ 
21 & 123758.7-683058 & 189.494717 & --68.516135 & 1.68 & 7.47 & 38$\pm$7 & 20$\pm$5 & 9.5 & 2.0$\pm$0.5 & -- \\ 
22 & 123842.1-681326 & 189.675256 & --68.224149 & 2.32 & 10.60 & 34$\pm$7 & 17$\pm$5 & 9.8 & 2.1$\pm$0.4 & + \\ 
23 & 123704.7-682331 & 189.269643 & --68.392221 & 0.97 & 5.74 & 28$\pm$6 & 4$\pm$2 & 6.9 & 3.4$\pm$0.3 & -- \\ 
24 & 123939.5-681858 & 189.914529 & --68.316227 & 2.38 & 9.67 & 28$\pm$6 & 16$\pm$4 & 7.0 & 1.7$\pm$0.4 & + \\ 
25 & 123918.7-681611 & 189.827710 & --68.269811 & 2.32 & 9.89 & 28$\pm$6 & 23$\pm$5 & 12 & 1.48$\pm$0.12 & + \\ 
26 & 123707.4-682334 & 189.280972 & --68.392931 & 0.96 & 5.49 & 26$\pm$5 & 11$\pm$4 & 6.0 & 2.6$\pm$0.7 & -- \\ 
27 & 123850.3-681639 & 189.709667 & --68.277553 & 1.44 & 7.96 & 25$\pm$5 & 21$\pm$5 & 6.0 & 1.15$\pm$0.18 & + \\ 
28 & 123716.3-682518 & 189.318042 & --68.421875 & 0.94 & 4.99 & 24$\pm$5 & 20$\pm$5 & 5.6 & 1.06$\pm$0.13 & + \\ 
29 & 123753.5-682000 & 189.472784 & --68.333349 & 0.50 & 3.76 & 24$\pm$5 & 17$\pm$4 & 6.8 & 1.51$\pm$0.17 & + \\ 
30 & 123836.9-682721 & 189.653872 & --68.455995 & 0.99 & 4.70 & 24$\pm$5 & 23$\pm$5 & 5.0 & 1.4$\pm$0.2 & + \\ 
31 & 123823.4-682820 & 189.597668 & --68.472407 & 1.13 & 5.03 & 24$\pm$5 & 18$\pm$4 & 6.1 & 1.40$\pm$0.14 & + \\ 
32 & 123716.3-682036 & 189.317855 & --68.343371 & 0.86 & 5.53 & 23$\pm$5 & 16$\pm$4 & 5.0 & 1.3$\pm$0.2 & -- \\ 
33 & 123804.7-682334 & 189.519709 & --68.392814 & 0.40 & 0.22 & 23$\pm$5 & 12$\pm$4 & 4.6 & 1.9$\pm$0.4 & -- \\ 
34 & 123858.4-681743 & 189.743326 & --68.295490 & 1.39 & 7.50 & 22$\pm$5 & 10$\pm$3 & 5.9 & 2.1$\pm$0.4 & -- \\ 
35 & 123732.9-682648 & 189.386958 & --68.446727 & 0.96 & 4.53 & 22$\pm$5 & 15$\pm$4 & 5.0 & 1.94$\pm$0.15 & -- \\ 
36 & 123656.3-682204 & 189.234675 & --68.367933 & 1.35 & 6.68 & 22$\pm$5 & 12$\pm$4 & 7.9 & 1.9$\pm$0.4 & + \\ 
37 & 123914.6-682316 & 189.810739 & --68.387795 & 1.30 & 6.22 & 20$\pm$5 & 12$\pm$4 & 4.5 & 1.9$\pm$0.5 & -- \\ 
38$\dagger$ & 123735.9-681430 & 189.399430 & --68.241885 & 2.47 & 9.48 & 20$\pm$5 & 12$\pm$4 & 6.1 & 1.8$\pm$0.2 & + \\ 
39 & 123805.6-682623 & 189.523310 & --68.439846 & 0.69 & 2.85 & 20$\pm$4 & 9$\pm$3 & 4.0 & 2.6$\pm$0.5 & -- \\ 
40 & 123715.7-682728 & 189.315302 & --68.457948 & 1.61 & 6.15 & 19$\pm$5 & 17$\pm$4 & 7.6 & 1.23$\pm$0.11 & + \\ 
41 & 123707.5-682443 & 189.281093 & --68.412219 & 1.22 & 5.61 & 19$\pm$5 & 8$\pm$3 & 4.3 & 2.5$\pm$0.5 & + \\ 
42 & 123846.8-682650 & 189.695026 & --68.447337 & 1.14 & 4.92 & 19$\pm$5 & 6$\pm$3 & 4.0 & 2.8$\pm$0.6 & -- \\ \hline
\multicolumn{11}{p {17cm}}{}                                             \\       
\multicolumn{11}{r}{continued on next page}\\
\end{tabular}
\end{center}
\end{table*}

%%%%%%%%%%%%%%%%%%%%%%%%%%%%%%%%%%%%%%%%%%%%%%%%%%%%%%%%%%%

\begin{table*}[h]
\begin{center}
\leavevmode
\begin{tabular}{lcccccccclc} \hline \hline              
(1) & (2) & (3) & (4) & (5) & (6) &(7) & (8) & (9) &(10) &  (11) \\
CX & CXOU\,J & $\alpha$ (J2000.0) & $\delta$ (J2000.0) & Error & $\theta$ & C$_{t,net}$ & C$_{s,net}$ & $F_{X,u}$ & $E_{50}$ & Optical Match    \\  
 & & (\arcdeg) & (\arcdeg) & (\arcsec) & (\arcmin) & & & (10$^{-15}$ erg cm$^{-2}$ s$^{-1}$) & (keV) & \\ \hline 
43 & 123709.4-682708 & 189.289121 & --68.452257 & 1.77 & 6.41 & 18$\pm$5 & 4$\pm$2 & 4.9 & 2.5$\pm$0.5 & -- \\ 
44$\dagger$ & 123640.6-682122 & 189.169352 & --68.356137 & 2.39 & 8.25 & 18$\pm$5 & 8$\pm$3 & 4.4 & 2.8$\pm$0.9 & + \\ 
45 & 123711.6-682036 & 189.298332 & --68.343498 & 1.10 & 5.89 & 18$\pm$4 & 7$\pm$3 & 3.8 & 2.1$\pm$0.7 & -- \\ 
46 & 123835.2-683046 & 189.646754 & --68.512900 & 2.90 & 7.68 & 18$\pm$5 & 12$\pm$4 & 4.2 & 1.2$\pm$0.5 & + \\ 
47 & 123751.1-682620 & 189.463038 & --68.438966 & 0.77 & 3.16 & 17$\pm$4 & 10$\pm$3 & 4.0 & 1.9$\pm$0.6 & + \\ 
48 & 123729.2-681706 & 189.371654 & --68.285207 & 1.49 & 7.32 & 17$\pm$4 & 12$\pm$4 & 4.7 & 1.6$\pm$0.4 & -- \\ 
49 & 123725.4-682443 & 189.355865 & --68.412196 & 0.86 & 4.01 & 16$\pm$4 & 15$\pm$4 & 3.5 & 1.26$\pm$0.20 & + \\ 
50 & 123808.9-681613 & 189.537120 & --68.270460 & 1.4 & 7.32 & 16$\pm$4 & 9$\pm$3 & 3.7 & 1.8$\pm$0.7 & -- \\ 
51 & 123759.1-682510 & 189.496064 & --68.419466 & 0.56 & 1.79 & 16$\pm$4 & 12$\pm$4 & 3.6 & 1.3$\pm$0.3 & + \\ 
52 & 123645.0-681926 & 189.187449 & --68.324003 & 2.73 & 8.61 & 16$\pm$5 & 7$\pm$3 & 3.7 & 2.2$\pm$0.6 & -- \\ 
53 & 123835.4-681447 & 189.647524 & --68.246512 & 2.69 & 9.13 & 15$\pm$5 & 10$\pm$4 & 4.1 & 1.8$\pm$0.8 & + \\ 
54 & 123810.7-682750 & 189.544666 & --68.464116 & 1.19 & 4.32 & 15$\pm$4 & 7$\pm$3 & 3.8 & 2.3$\pm$0.5 & + \\ 
55$\dagger$ & 123928.7-682454 & 189.869435 & --68.415071 & 2.58 & 7.63 & 15$\pm$5 & 7$\pm$3 & 3.4 & 1$\pm$2 & + \\ 
56 & 123741.3-682130 & 189.422082 & --68.358334 & 0.55 & 3.13 & 15$\pm$4 & 9$\pm$3 & 3.4 & 1.6$\pm$0.4 & -- \\ 
57$\dagger$ & 123913.9-681834 & 189.807786 & --68.309638 & 2.15 & 7.91 & 15$\pm$4 & 7$\pm$3 & 3.7 & 2.2$\pm$0.5 & + \\ 
58 & 123912.6-682737 & 189.802665 & --68.460385 & 2.64 & 7.28 & 15$\pm$5 & 15$\pm$4 & 3.5 & 1.4$\pm$0.1 & + \\ 
59 & 123752.8-682559 & 189.469961 & --68.433066 & 0.74 & 2.77 & 15$\pm$4 & 15$\pm$4 & 3.8 & 0.95$\pm$0.06 & + \\ 
60 & 123656.6-682452 & 189.235923 & --68.414656 & 1.98 & 6.62 & 14$\pm$4 & 7$\pm$3 & 3.7 & 2.4$\pm$0.7 & -- \\ 
61 & 123800.5-682710 & 189.502237 & --68.452840 & 1.00 & 3.68 & 14$\pm$4 & 7$\pm$3 & 3.3 & 2.3$\pm$0.8 & + \\ 
62$\dagger$ & 123911.1-681705 & 189.796169 & --68.284874 & 2.76 & 8.75 & 14$\pm$5 & 5$\pm$3 & 3.4 & 2$\pm$1 & + \\ 
63$\dagger$ & 123926.8-682508 & 189.861487 & --68.418971 & 2.66 & 7.50 & 14$\pm$4 & 12$\pm$4 & 3.4 & 1.2$\pm$0.2 & + \\ 
64 & 123708.8-682321 & 189.286700 & --68.389211 & 1.27 & 5.37 & 14$\pm$4 & 11$\pm$3 & 3.0 & 1.3$\pm$0.3 & + \\ 
65 & 123802.7-683000 & 189.511184 & --68.500163 & 2.50 & 6.48 & 14$\pm$4 & 7$\pm$3 & 3.2 & 1.8$\pm$0.9 & + \\ 
66$\dagger$ & 123724.4-682854 & 189.351711 & --68.481709 & 2.53 & 6.64 & 13$\pm$4 & 8$\pm$3 & 3.6 & 2$\pm$1 & -- \\ 
67 & 123908.7-682405 & 189.786450 & --68.401525 & 1.51 & 5.70 & 13$\pm$4 & 8$\pm$3 & 3.1 & 1.9$\pm$0.3 & + \\ 
68 & 123732.9-682254 & 189.387088 & --68.381922 & 0.67 & 3.21 & 13$\pm$4 & 9$\pm$3 & 2.7 & 1.0$\pm$0.3 & + \\ 
69$\dagger$ & 123933.4-681758 & 189.889151 & --68.299714 & 4.41 & 9.71 & 12$\pm$5 & 5$\pm$3 & 3.1 & 2$\pm$1 & + \\ 
70 & 123758.4-682301 & 189.493504 & --68.383764 & 0.44 & 0.95 & 12$\pm$4 & 9$\pm$3 & 3.7 & 1.6$\pm$0.3 & + \\ 
71 & 123819.9-681908 & 189.582922 & --68.319011 & 0.76 & 4.56 & 12$\pm$4 & 7$\pm$3 & 2.4 & 1.8$\pm$0.8 & -- \\ 
72 & 123858.3-682703 & 189.742924 & --68.450945 & 2.07 & 5.88 & 11$\pm$4 & 4$\pm$2 & 2.5 & 2.1$\pm$0.4 & -- \\ 
73 & 123859.9-682307 & 189.749388 & --68.385393 & 1.18 & 4.88 & 11$\pm$4 & 11$\pm$3 & 2.3 & 1.05$\pm$0.18 & + \\ 
74 & 123744.2-682828 & 189.434117 & --68.474621 & 1.95 & 5.36 & 11$\pm$4 & 11$\pm$4 & 2.6 & 1.1$\pm$0.3 & + \\ 
75 & 123833.4-681621 & 189.639295 & --68.272733 & 2.08 & 7.58 & 11$\pm$4 & 7$\pm$3 & 2.6 & 1.6$\pm$0.9 & -- \\ 
76 & 123839.4-682609 & 189.664144 & --68.436102 & 1.14 & 3.97 & 11$\pm$4 & 5$\pm$2 & 2.3 & 2.4$\pm$0.5 & -- \\ 
77$\dagger$ & 123920.3-681804 & 189.834519 & --68.301311 & 3.40 & 8.69 & 11$\pm$4 & 4$\pm$2 & 2.6 & 2.7$\pm$0.9 & + \\ 
78 & 123759.8-682303 & 189.499214 & --68.384224 & 0.45 & 0.83 & 11$\pm$3 & 6$\pm$3 & 6.0 & 2.0$\pm$0.4 & + \\ 
79 & 123755.2-682514 & 189.479947 & --68.420605 & 0.67 & 2.02 & 11$\pm$3 & 7$\pm$3 & 2.3 & 1.5$\pm$0.4 & -- \\ 
80$\dagger$ & 123723.4-682745 & 189.347313 & --68.462565 & 2.21 & 5.82 & 11$\pm$4 & 6$\pm$3 & 2.6 & 2$\pm$1 & + \\ 
81 & 123809.9-682032 & 189.541149 & --68.342477 & 0.53 & 3.00 & 11$\pm$3 & 10$\pm$3 & 2.5 & 1.2$\pm$0.2 & + \\ 
82 & 123811.7-682522 & 189.548553 & --68.423049 & 0.67 & 1.89 & 11$\pm$3 & 9$\pm$3 & 2.2 & 1.5$\pm$0.2 & + \\ 
83$\dagger$ & 123755.0-681542 & 189.479307 & --68.261917 & 2.39 & 7.91 & 11$\pm$4 & 2$\pm$2 & 2.7 & 3.3$\pm$0.5 & + \\ 
84 & 123813.6-682637 & 189.556813 & --68.443847 & 0.99 & 3.15 & 10$\pm$3 & 3$\pm$2 & 2.2 & 2.1$\pm$0.6 & -- \\ \hline
\multicolumn{11}{p {17cm}}{}                                             \\       
\multicolumn{11}{r}{continued on next page}\\
\end{tabular}
\end{center}
\end{table*}

%%%%%%%%%%%%%%%%%%%%%%%%%%%%%%%%%%%%%%%%%%%%%%%%%%%%%%%%%%%%

\begin{table*}[h]
\begin{center}
\begin{tabular}{lcccccccclc} \hline \hline              
(1) & (2) & (3) & (4) & (5) & (6) &(7) & (8) & (9) &(10) &  (11) \\
CX & CXOU\,J & $\alpha$ (J2000.0) & $\delta$ (J2000.0) & Error & $\theta$ & C$_{t,net}$ & C$_{s,net}$ & $F_{X,u}$ & $E_{50}$ & Optical Match    \\ 
 & & (\arcdeg) & (\arcdeg) & (\arcsec) & (\arcmin) & & & (10$^{-15}$ erg cm$^{-2}$ s$^{-1}$) & (keV) & \\ \hline 
85 & 123815.4-682427 & 189.564219 & --68.407599 & 0.57 & 1.19 & 10$\pm$3 & 7$\pm$3 & 2.0 & 1.2$\pm$0.5 & + \\ 
86 & 123816.7-682436 & 189.569630 & --68.410056 & 0.59 & 1.38 & 10$\pm$3 & 6$\pm$3 & 2.0 & 1.4$\pm$0.7 & + \\ 
87 & 123828.2-682116 & 189.617623 & --68.354598 & 0.58 & 2.99 & 10$\pm$3 & 8$\pm$3 & 2.0 & 1.3$\pm$0.3 & + \\ 
88 & 123823.4-681706 & 189.597632 & --68.285277 & 1.63 & 6.60 & 10$\pm$3 & 10$\pm$3 & 2.2 & 1.03$\pm$0.16 & + \\ 
89 & 123757.3-682502 & 189.488645 & --68.417264 & 0.68 & 1.75 & 9$\pm$3 & 7$\pm$3 & 2.1 & 1.1$\pm$0.4 & + \\ 
90 & 123818.0-682417 & 189.574953 & --68.404962 & 0.58 & 1.26 & 9$\pm$3 & 6$\pm$3 & 1.8 & 1.5$\pm$0.5 & -- \\ 
91$\dagger$ & 123844.8-682606 & 189.686473 & --68.435269 & 1.44 & 4.32 & 9$\pm$3 & 6$\pm$3 & 1.8 & 1.3$\pm$0.9 & + \\ 
92 & 123630.8-682249 & 189.128208 & --68.380359 & 5.51 & 8.90 & 9$\pm$4 & 9$\pm$4 & 2.6 & 1.39$\pm$0.15 & + \\ 
93 & 123731.1-682128 & 189.379664 & --68.357778 & 0.86 & 3.91 & 9$\pm$3 & 5$\pm$2 & 1.9 & 1$\pm$1 & + \\ 
94 & 123800.6-682159 & 189.502590 & --68.366532 & 0.47 & 1.66 & 9$\pm$3 & 9$\pm$3 & 1.8 & 0.99$\pm$0.10 & + \\ 
95 & 123832.8-682202 & 189.636665 & --68.367275 & 0.63 & 2.81 & 9$\pm$3 & 9$\pm$3 & 1.7 & 1.12$\pm$0.16 & + \\ 
96$\dagger$ & 123918.9-682558 & 189.828694 & --68.432881 & 3.44 & 7.04 & 9$\pm$4 & 2$\pm$2 & 2.2 & 2.6$\pm$0.3 & + \\ 
97$\dagger$ & 123812.1-681712 & 189.550429 & --68.286827 & 1.60 & 6.35 & 9$\pm$3 & 8$\pm$3 & 2.0 & 1.13$\pm$0.18 & + \\ 
98 & 123828.0-682442 & 189.616733 & --68.411870 & 0.74 & 2.25 & 9$\pm$3 & 6$\pm$2 & 1.8 & 1.4$\pm$0.7 & -- \\ 
99 & 123755.7-682607 & 189.482014 & --68.435438 & 0.97 & 2.79 & 9$\pm$3 & 6$\pm$2 & 2.1 & 1.5$\pm$0.6 & -- \\ 
100 & 123835.4-682622 & 189.647542 & --68.439534 & 1.36 & 3.85 & 8$\pm$3 & 9$\pm$3 & 1.7 & 0.93$\pm$0.08 & -- \\ 
101 & 123839.9-682811 & 189.666443 & --68.469832 & 2.61 & 5.54 & 8$\pm$3 & 5$\pm$2 & 1.7 & 1.1$\pm$0.7 & -- \\ 
102 & 123810.6-682104 & 189.544130 & --68.351119 & 0.53 & 2.50 & 8$\pm$3 & 2$\pm$2 & 1.8 & 2.9$\pm$0.5 & -- \\ 
103 & 123823.8-682330 & 189.599256 & --68.391940 & 0.58 & 1.54 & 8$\pm$3 & 5$\pm$2 & 1.5 & 1.7$\pm$0.7 & + \\ 
104$\dagger$ & 123833.7-682011 & 189.640396 & --68.336559 & 0.88 & 4.15 & 8$\pm$3 & 4$\pm$2 & 1.7 & 2.0$\pm$0.4 & -- \\ 
105 & 123841.7-682104 & 189.673821 & --68.351302 & 0.92 & 4.03 & 8$\pm$3 & 7$\pm$3 & 1.6 & 1.5$\pm$0.2 & -- \\ 
106 & 123836.8-682013 & 189.653477 & --68.337185 & 0.94 & 4.30 & 8$\pm$3 & 5$\pm$2 & 1.7 & 1.8$\pm$0.3 & -- \\ 
107 & 123839.3-682014 & 189.663567 & --68.337314 & 1.00 & 4.44 & 8$\pm$3 & 2$\pm$2 & 1.7 & 2.7$\pm$0.8 & -- \\ 
108$\dagger$ & 123806.6-682616 & 189.527489 & --68.437998 & 1.05 & 2.74 & 8$\pm$3 & 3$\pm$2 & 1.6 & 3.3$\pm$0.8 & -- \\ 
109$\dagger$ & 123811.4-681308 & 189.547534 & --68.219135 & 7.78 & 10.40 & 7$\pm$4 & $<$2.5 & 2.1 & 5$\pm$1 & + \\ 
110 & 123815.9-682134 & 189.566252 & --68.359664 & 0.53 & 2.12 & 7$\pm$3 & 2$\pm$2 & 3.8 & 2.2$\pm$0.8 & -- \\ 
111 & 123738.9-682118 & 189.412214 & --68.355234 & 0.80 & 3.42 & 7$\pm$3 & $<$1.1 & 1.6 & 4.8$\pm$0.5 & -- \\ 
112 & 123800.2-682511 & 189.500897 & --68.419799 & 0.78 & 1.76 & 7$\pm$3 & 3$\pm$2 & 1.6 & 2.1$\pm$0.7 & + \\ 
113 & 123745.9-682109 & 189.441337 & --68.352636 & 0.69 & 3.08 & 7$\pm$3 & 3$\pm$2 & 1.8 & 2.1$\pm$0.8 & -- \\ 
114 & 123826.9-681859 & 189.611891 & --68.316430 & 1.16 & 4.91 & 7$\pm$3 & 6$\pm$3 & 1.5 & 1.1$\pm$0.4 & + \\ 
115 & 123826.1-682534 & 189.608694 & --68.426193 & 1.04 & 2.68 & 6$\pm$3 & 5$\pm$2 & 1.3 & 1$\pm$1 & + \\  
116 & 123841.3-682447 & 189.672185 & --68.413140 & 1.19 & 3.39 & 6$\pm$3 & 4$\pm$2 & 1.3 & 2$\pm$1 & -- \\ 
117$\dagger$ & 123718.3-681822 & 189.326262 & --68.306383 & 2.86 & 6.84 & 6$\pm$3 & 2$\pm$2 & 1.3 & 3$\pm$2 & -- \\ 
118 & 123812.7-682301 & 189.552882 & --68.383798 & 0.53 & 0.73 & 6$\pm$3 & 2$\pm$2 & 1.2 & 3.0$\pm$0.9 & -- \\ 
119$\dagger$ & 123714.9-682018 & 189.311995 & --68.338503 & 2.18 & 5.80 & 6$\pm$3 & 7$\pm$3 & 1.3 & 0.9$\pm$0.2 & + \\ 
120$\dagger$ & 123829.0-681920 & 189.620718 & --68.322330 & 1.25 & 4.66 & 5$\pm$2 & 5$\pm$2 & 1.1 & 1.2$\pm$1.1 & + \\ 
121$\dagger$ & 123809.3-681438 & 189.538731 & --68.244009 & 6.38 & 8.90 & 5$\pm$3 & 6$\pm$3 & 1.2 & 1.05$\pm$0.06 & + \\ 
122 & 123832.6-682324 & 189.635758 & --68.390217 & 0.85 & 2.35 & 5$\pm$2 & 0. & 2.2 & 3.7$\pm$0.6 & -- \\ 
123 & 123759.7-682349 & 189.498816 & --68.397053 & 0.65 & 0.73 & 5$\pm$2 & 3$\pm$2 & 1.0 & 1$\pm$1 & + \\ 
124 & 123805.9-682255 & 189.524767 & --68.382160 & 0.56 & 0.62 & 5$\pm$2 & 4$\pm$2 & 0.97 & 1.1$\pm$0.4 & + \\ 
125 & 123808.4-681937 & 189.535049 & --68.327083 & 0.95 & 3.92 & 5$\pm$2 & 4$\pm$2 & 0.99 & 1$\pm$1 & + \\ 
126 & 123811.6-682410 & 189.548339 & --68.402850 & 0.69 & 0.75 & 5$\pm$2 & 5$\pm$2 & 0.97 & 1.2$\pm$0.3 & + \\ \hline
\multicolumn{11}{p {17cm}}{}                                             \\       
\multicolumn{11}{r}{continued on next page}\\
\end{tabular}
\end{center}
\end{table*}

%%%%%%%%%%%%%%%%%%%%%%%%%%%%%%%%%%%%%%%%%%%%%%%%%%%%%%%%%%%%

\begin{table*}[h]
\begin{center}
\begin{tabular}{lcccccccclc} \hline \hline              
(1) & (2) & (3) & (4) & (5) & (6) &(7) & (8) & (9) &(10) &  (11) \\
CX & CXOU\,J & $\alpha$ (J2000.0) & $\delta$ (J2000.0) & Error & $\theta$ & C$_{t,net}$ & C$_{s,net}$ & $F_{X,u}$ & $E_{50}$ & Optical Match    \\ 
 & & (\arcdeg) & (\arcdeg) & (\arcsec) & (\arcmin) & & & (10$^{-15}$ erg cm$^{-2}$ s$^{-1}$) & (keV) & \\ \hline 
127$\dagger$ & 123919.1-682346 & 189.829698 & --68.396355 & 4.42 & 6.64 & 5$\pm$3 & 6$\pm$3 & 1.3 & 1.0$\pm$0.2 & + \\ 
128$\dagger$ & 123727.2-682316 & 189.363451 & --68.387783 & 1.36 & 3.68 & 5$\pm$2 & 5$\pm$2 & 1.1 & 0.9$\pm$0.3 & + \\ 
129 & 123743.7-682518 & 189.432199 & --68.421940 & 1.26 & 2.79 & 5$\pm$2 & 5$\pm$2 & 1.1 & 1.2$\pm$0.2 & + \\ 
130 & 123814.3-682002 & 189.559538 & --68.334078 & 0.86 & 3.56 & 5$\pm$2 & 5$\pm$2 & 0.97 & 0.89$\pm$0.13 & + \\ 
131$\dagger$ & 123646.3-682528 & 189.192806 & --68.424563 & 7.43 & 7.68 & 5$\pm$3 & $<$1.1 & 1.3 & 5.9$\pm$0.7 & + \\ 
132 & 123845.3-682336 & 189.688697 & --68.393346 & 1.34 & 3.52 & 5$\pm$2 & 5$\pm$2 & 1.1 & 0.87$\pm$0.10 & + \\ 
133$\dagger$ & 123744.6-682543 & 189.436025 & --68.428709 & 1.45 & 3.00 & 5$\pm$2 & 5$\pm$2 & 1.2 & 1.6$\pm$0.2 & + \\ 
134 & 123822.2-682557 & 189.592308 & --68.432742 & 1.39 & 2.79 & 5$\pm$2 & 5$\pm$2 & 0.95 & 0.9$\pm$0.2 & + \\ 
135$\dagger$ & 123731.0-681841 & 189.379262 & --68.311663 & 2.37 & 5.87 & 4$\pm$2 & 4$\pm$2 & 1.8 & 1.7$\pm$0.8 & + \\ 
136 & 123751.4-682234 & 189.464318 & --68.376236 & 0.72 & 1.74 & 4$\pm$2 & 4$\pm$2 & 0.94 & 1.09$\pm$0.13 & + \\ 
137 & 123758.9-682407 & 189.495337 & --68.402109 & 0.79 & 0.95 & 4$\pm$2 & 4$\pm$2 & 0.80 & 1.58$\pm$0.18 & + \\ 
138$\dagger$ & 123800.1-682234 & 189.500230 & --68.376235 & 0.63 & 1.16 & 4$\pm$2 & 3$\pm$2 & 1.1 & 1.5$\pm$0.4 & + \\ 
139$\dagger$ & 123805.5-682100 & 189.523107 & --68.350180 & 0.72 & 2.53 & 4$\pm$2 & 4$\pm$2 & 1.1 & 1.22$\pm$0.18 & + \\ 
140$\dagger$ & 123812.4-682441 & 189.551614 & --68.411460 & 0.90 & 1.25 & 4$\pm$2 & 2$\pm$2 & 0.86 & 2$\pm$1 & -- \\ 
141$\dagger$ & 123735.4-682040 & 189.397562 & --68.344524 & 1.39 & 4.09 & 4$\pm$2 & $<$1.1 & 0.94 & 4$\pm$1 & -- \\ 
142 & 123805.2-682531 & 189.521504 & --68.425483 & 1.21 & 1.99 & 4$\pm$2 & 4$\pm$2 & 0.89 & 1.5$\pm$0.3 & + \\ 
143 & 123806.8-681914 & 189.528172 & --68.320569 & 1.29 & 4.31 & 4$\pm$2 & 4$\pm$2 & 0.77 & 0.9$\pm$0.3 & + \\ 
144$\dagger$ & 123842.4-682204 & 189.676583 & --68.367882 & 1.33 & 3.57 & 4$\pm$2 & 4$\pm$2 & 0.79 & 0.7$\pm$0.3 & + \\ 
145$\dagger$ & 123828.5-682624 & 189.618784 & --68.440128 & 2.17 & 3.48 & 4$\pm$2 & 1$\pm$1 & 0.75 & 2.8$\pm$0.7 & -- \\ 
146$\dagger$ & 123819.5-681830 & 189.581264 & --68.308352 & 1.97 & 5.17 & 4$\pm$2 & 1$\pm$1 & 0.72 & 3$\pm$2 & -- \\ 
147$\dagger$ & 123816.4-682213 & 189.568335 & --68.370356 & 0.74 & 1.58 & 3$\pm$2 & 3$\pm$2 & 1.6 & 0.7$\pm$0.3 & + \\ 
148$\dagger$ & 123800.9-682226 & 189.503632 & --68.373923 & 0.73 & 1.25 & 3$\pm$2 & 3$\pm$2 & 0.73 & 1.19$\pm$0.12 & + \\ 
149$\dagger$ & 123811.8-681934 & 189.549265 & --68.326171 & 1.37 & 4.00 & 3$\pm$2 & 3$\pm$2 & 0.58 & 1.1$\pm$0.3 & + \\ 
150$\dagger$ & 123813.7-682608 & 189.557289 & --68.435640 & 1.96 & 2.67 & 3$\pm$2 & $<$1.1 & 0.72 & 3.6$\pm$0.5 & + \\ 
151$\dagger$ & 123815.3-681941 & 189.563823 & --68.328264 & 1.37 & 3.92 & 3$\pm$2 & 3$\pm$2 & 0.57 & 1.69$\pm$0.17 & + \\ \hline

\multicolumn{11}{p {17cm}}{}                                             \\       
\multicolumn{11}{p {17cm}}{Col.\ (1): X-ray catalogue sequence number, sorted by net X-ray counts (0.3--7~keV). Sources that were detected by {\tt wavdetect} using a {\tt sigthresh} of $10^{-6}$ but not with a {\tt sigthresh} of $10^{-7}$ have been flagged with a $\dagger$.}\\
\multicolumn{11}{p {17cm}}{Col.\ (2): IAU designated source name. }                   \\
\multicolumn{11}{p {17cm}}{Cols.\ (3) and (4): Right ascension and declination (in decimal degrees) for epoch J2000.0.} \\
\multicolumn{11}{p {17cm}}{Col.\ (5): 95\% confidence radius on {\tt wavdetect} X-ray source position in arcseconds.}  \\
\multicolumn{11}{p {17cm}}{Col.\ (6): Angular offset from our derived cluster centre ($\alpha_{2000} = 12^{\rm h}38^{\rm m}07\fs1$, $\delta_{2000} =
-68^{\circ}23\arcmin33\arcsec$) in arcminutes.}\\
\multicolumn{11}{p {17cm}}{Col.\ (7): Net counts extracted in the total energy band (0.3--7~keV) with 1-$\sigma$ errors.}\\
\multicolumn{11}{p {17cm}}{Col.\ (8): Net counts extracted in the soft energy band (0.3--2~keV) with 1-$\sigma$ errors. For sources CX\,109, CX\,111, CX\,131, CX\,141 and CX\,150, only 1-$\sigma$ upper limits are given.}\\
\multicolumn{11}{p {17cm}}{Col.\ (9): Unabsorbed X-ray flux in the 0.3--7~keV energy band for a 2~keV MeKaL model and neutral hydrogen column of 1.9$\times$10$^{21}$cm$^{-2}$.}\\
\multicolumn{11}{p {17cm}}{Col.\ (10): Median energy $E_{50}$ in keV with 1-$\sigma$ errors.}\\
\multicolumn{11}{p {17cm}}{Col.\ (11): Information about presence (+) or absence (--) of optical counterpart (details in Table \ref{tab2}).}\\
\end{tabular}
\end{center}
\end{table*}

%
% Table 2
%

\begin{table*}[h]
\caption{Optical Counterpart Properties}
\label{tab2}
\begin{center}
\begin{tabular}{clclccccccc} \hline \hline              
(1) & (2) & (3) & (4) & (5) & (6) &(7) & (8) & (9) &(10) & (11)  \\
CX & OID & Dox & $V$ & $B-V$ & Var & P & Variable Type & $L_{X,u}$ &log($F_{X}$/$F_{V}$)$_{u}$ & Class \\
 & & (\arcsec) &  & & & (days) & & (10$^{30}$ erg s$^{-1}$) & &  \\ \hline
4 & 25708 & 0.23 & 21.65 & 1.27 & -- & -- & -- &  25.5 & 0.21 & CV? \\ 
7 & 108-058885 & 0.34 & 11.42 & 1.06 & -- & -- & -- &  21.3 & --3.96 & NM \\ 
8 & 22375 & 0.18 & 21.10 & 0.77 & -- & -- & -- &  16.9 & --0.19 & CV? \\ 
9 & 15550 & 0.45 & 15.458 & 1.150 & -- & -- & -- & 14.5 & --2.51 & YSS \\ 
10 & 14106 & 0.11 & 20.40 & 1.88 & -- & -- & -- & 11.3 & --0.64 & NM \\ 
12 & 15171 & 0.62 & 16.739 & 1.382 & -- & -- & -- & 12.3 & --2.07 & SSG \\ 
13 & 15326 & 0.30 & 16.042 & 0.904 & -- & -- & -- & 9.25 & --2.47 & AB \\ 
14 & 31878 & 1.91 & 20.73 & 1.68 & -- & -- & -- & 10.4 & --0.55 & Unc \\ 
 & 31966 & 1.96 & 19.24 & 1.37 & -- & -- & -- & 10.4 & --1.14 & Unc \\ 
15 & 29067 & 1.79 & 18.903 & 1.617 & -- & -- & -- & 12.5 & --1.20 & NM \\ 
17 & 35116$^{c}$ & 1.08 & 22.9$^{b}$ & -- & -- & -- & -- & 9.85 & 0.05 & Unc\\ 
 & 35119$^{c}$ & 1.03 & 22.4 & -- & -- & -- & -- & 9.85 & 0.09 & Unc \\ 
18 & 18485 & 0.12 & 14.361 & 0.826 & V45 & 2.11? & EB(BSS) & 8.05 & --3.20 & BSS \\ 
20 & 19384 & 1.40 & 21.79 & 0.6 & -- & -- & -- & 9.07 & --0.18 & CV? \\ 
22 & 35113 & 1.51 & 23.3$^{b}$ & -- & -- & -- & -- & 8.49 & 0.15 & CV? \\ 
24 & 30048 & 2.37 & 20.37 & 1.03 & -- & -- & -- & 6.08 & --0.92 & CV? \\ 
 & 35114$^{a}$ & 2.39 & 23.8$^{b}$ & -- & -- & -- & -- & 6.08 & 0.20 & CV? \\ 
25 & 27968 & 1.12 & 16.381 & 1.204 & -- & -- & -- & 10.3 & --2.29 & AB? \\ 
27 & 23901 & 0.86 & 16.452 & 1.043 & V42 & 0.7029 & EW &  5.29 & --2.55 & AB \\ 
28 & 9861$^{c}$ & 1.29 & 16.164 & 1.268 & -- & -- & -- & 4.92 & --2.70 & AB \\ 
29 & 35093 & 0.32 & 22.5 & -- & -- & -- & -- & 6.04 & --0.09 & Unc \\ 
30 & 108-058907 & 0.36 & 13.11 & 0.53 & -- & -- & -- & 4.43 & --3.97 & NM \\ 
31 & 19725 & 0.27 & 17.107 & 1.324 & -- & -- & -- & 5.39 & --2.28 & SSG \\ 
36 & 35102 & 1.38 & 22.1 & -- & -- & -- & -- & 7.01 & --0.17 & Unc \\ 
38 & 12668 & 1.28 & 20.56 & 1.31 & -- & -- & -- & 5.36 & --0.90 & CV? \\ 
 & 12714 & 1.75 & 22.0 & 1.1 & -- & -- & -- & 5.36 & --0.31 & CV? \\ 
40 & 108-058509 & 0.12 & 13.52 & 0.53 & -- & -- & -- & 6.65 & --3.62 & NM \\ 
41 & 8654 & 0.81 & 20.15 & 1.34 & -- & -- & -- & 3.80 & --1.22 & AB? \\ 
44 & 4648 & 1.18 & 21.67 & 1.28 & -- & -- & -- & 3.89 & --0.60 & CV? \\ 
46 & 21648 & 1.73 & 19.57 & 1.47 & -- & -- & -- & 3.71 & --1.46 & AB \\ 
 & 21731 & 2.74 & 19.00 & 1.45 & -- & -- & -- & 3.71 & --1.69 & AB \\ 
47 & 15107 & 0.73 & 19.83 & 1.43 & -- & -- & -- & 3.53 & --1.37 & AB \\ 
49 & 11157 & 0.36 & 15.899 & 1.014 & -- & -- & -- & 3.09 & --3.01 & AB \\ 
51 & 15544 & 0.30 & 17.558 & 1.116 & -- & -- & -- & 3.20 & --2.33 & AB \\ 
53 & 35086 & 1.65 & 21.51 & 2.0 & -- & -- & -- & 3.60 & --0.70 & NM \\ 
54 & 17413 & 1.26 & 21.20 & 1.4 & -- & -- & -- & 3.39 & --0.85 & CV? \\ 
55 & 35094 & 1.06 & 17.623 & 0.8$^{e}$ & -- & -- & -- & 2.97 & --2.33 & NM \\ 
57 & 27290 & 0.43 & 17.563 & 1.027 & -- & -- & -- & 3.22 & --2.32 & AB? \\ 
58 & 27108 & 0.38 & 17.744 & 1.422 & -- & -- & -- & 3.07 & --2.27 & SSG \\ 
59 & 35095 & 0.39 & 13.75 & -- & -- & -- & -- & 3.33 & --3.83 & NM \\ 
61 & 15835$^{c}$ & 1.27 & 18.38 & 1.29 & -- & -- & -- & 2.88 & --2.04 & AB \\ \hline
\multicolumn{11}{p {17cm}}{}                                             \\       
\multicolumn{11}{r}{continued on next page}\\
\end{tabular}
\end{center}
\end{table*}

\begin{table*}[h]
\begin{center}
\begin{tabular}{clclccccccc} \hline \hline              
(1) & (2) & (3) & (4) & (5) & (6) &(7) & (8) & (9) &(10) & (11) \\
CX & OID & Dox & $V$ & $B-V$ & Var & P & Variable Type & $L_{X,u}$ &log($F_{X}$/$F_{V}$)$_{u}$ & Class \\
 & & (\arcsec) & & & & (days) & & (10$^{30}$ erg s$^{-1}$) & &  \\ \hline
62 & 26831 & 2.77 & 16.550 & 1.037 & -- & -- & -- & 2.93 & --2.77 & AB? \\ 
 & 26933$^{c}$ & 2.85 & 18.123 & 1.084 & -- & -- & -- & 2.93 & --2.14 & AB? \\ 
63 & 35096 & 1.75 & 19.95 & -- & -- & -- & -- & 2.98 & --1.40 & Unc \\ 
64 & 35103 & 0.90 & 19.81 & -- & -- & -- & -- & 2.64 & --1.51 & Unc \\ 
 & 35104 & 0.80 & 19.68 & -- & -- & -- & -- & 2.64 & --1.56 & Unc \\ 
65 & 35117 & 1.31 & 22.8$^{b}$ & -- & -- & -- & -- & 2.81 & --0.52 & Unc \\ 
 & 35118$^{a}$ & 0.41 & 19.8 & 1.2 & -- & -- & -- & 2.81 & --1.47 & Unc \\ 
67 & 26543 & 0.10 & 16.520 & 0.792 & -- & -- & -- & 2.78 & --2.80 & BSS \\ 
68 & 12250 & 0.15 & 20.27 & 1.88 & V20 & 0.57712 & EA & 2.37 & --1.37 & NM-AB \\ 
69 & 35105 & 3.62 & 22.7 & -- & -- & -- & -- & 2.71 & --0.34 & Unc \\ 
70 & 15446 & 0.36 & 22.3 & 0.9 & -- & -- & -- & 3.30 & --0.41 & CV? \\ 
73 & 25282 & 0.57 & 14.028 & 0.893 & -- & -- & -- & 2.07 & --3.93 & BSS \\
74 & 13995 & 0.39 & 14.606 & 0.876 & -- & -- & -- & 2.28 & --3.65 & BSS \\  
77 & 28170 & 1.58 & 20.97 & 1.9 & -- & -- & -- & 2.29 & --1.11 & NM \\ 
78 & 15662 & 0.52 & 17.134 & 1.104 & -- & -- & -- & 5.34 & --2.27 & AB \\ 
80 & 10786 & 2.01 & 19.35 & 1.18 & -- & -- & -- & 2.31 & --1.75 & AB? \\ 
81 & 17291$^{a}$ & 0.43 & 17.41 & 1.08 & -- & -- & -- & 2.18 & --2.56 & AB \\ 
82 & 17610 & 0.21 & 19.19 & 1.26 & -- & -- & -- & 1.96 & --1.89 & AB \\ 
83 & 35097 & 1.34 & 22.4 & -- & -- & -- & -- & 2.37 & --0.50 & Unc \\ 
 & 35098 & 2.40 & 21.13 & -- & -- & -- & -- & 2.37 & --1.03 & Unc \\ 
 & 35099 & 1.95 & 20.53 & -- & -- & -- & -- & 2.37 & --1.27 & Unc \\ 
85 & 18293 & 0.10 & 18.83 & 1.23 & -- & -- & -- & 1.78 & --2.07 & AB \\ 
86 & 18530 & 0.11 & 15.781 & 0.824 & V12 & 1.4226 & EA(BSS) & 1.78 & --3.29 & BSS \\ 
87 & 20540 & 0.16 & 19.50 & 1.48 & -- & -- & -- & 1.73 & --1.82 & AB \\ 
88 & 19728 & 0.53 & 19.90 & 1.49 & -- & -- & -- & 1.98 & --1.60 & AB \\ 
89 & 35100 & 0.33 & 18.9$^{d}$ & 1.18$^{d}$ & V11 & 0.5405 & EB/EA & 1.82 & --2.04 & AB \\ 
91 & 35090 & 1.48 & 20.63 & 2.0 & -- & -- & -- & 1.63 & --1.39 & NM \\ 
92 & 3202 & 2.83 & 20.90 & 0.97 & -- & -- & -- & 2.26 & --1.14 & Unc \\ 
 & 3229 & 2.34 & 18.298 & 1.18 & -- & -- & -- & 2.26 & --2.18 & Unc \\ 
 & 3251 & 0.77 & 20.37 & 1.27 & -- & -- & -- & 2.26 & --1.35 & Unc \\ 
 & 35088 & 3.72 & 21.41 & 1.5 & -- & -- & -- & 2.26 & --0.94 & Unc \\ 
 & 35089 & 3.28 & 21.63 & 1.6 & -- & -- & -- & 2.26 & --0.85 & Unc \\ 
93 & 12003 & 0.31 & 17.078 & 1.115 & V30 & 0.35132 & EW & 1.72 & --2.79 & AB \\ 
94 & 15804 & 0.09 & 16.440 & 1.107 & -- & -- & -- & 1.56 & --3.09 & AB \\ 
95 & 21300 & 0.31 & 13.744 & 0.72 & -- & -- & -- & 1.55 & --4.17 & NM \\ 
96 & 27916 & 2.28 & 19.93 & 1.28 & -- & -- & -- & 1.96 & --1.59 & Unc \\ 
 & 28023 & 1.49 & 21.06 & 0.96 & -- & -- & -- & 1.96 & --1.14 & Unc \\ 
97 & 17685 & 0.45 & 18.221 & 1.165 & V38 & 1.31? & EA & 1.73 & --2.33 & AB \\ 
100 & 108-058898 & 0.78 & 12.372 & 0.675 & -- & -- & -- & 1.50 & --4.73 & NM \\ 
103 & 19794 & 0.19 & 18.714 & 1.120 & -- & -- & -- & 1.37 & --2.23 & AB \\  \hline
\multicolumn{11}{p {17cm}}{}                                             \\       
\multicolumn{11}{r}{continued on next page}\\
\end{tabular}
\end{center}
\end{table*}

\begin{table*}[h]
\begin{center}
\begin{tabular}{clclccccccc} \hline \hline              
(1) & (2) & (3) & (4) & (5) & (6) &(7) & (8) & (9) &(10) & (11) \\
CX & OID & Dox & $V$ & $B-V$ & Var & P & Variable Type & $L_{X,u}$ &log($F_{X}$/$F_{V}$)$_{u}$ & Class \\
 & & (\arcsec) & & & & (days) & & (10$^{30}$ erg s$^{-1}$) & &  \\ \hline
109 & 17333 & 7.61 & 19.88 & 1.32 & -- & -- & -- & 1.86 & --1.64 & Unc \\ 
 & 17434 & 7.47 & 15.175 & 1.384 & -- & -- & -- & 1.86 & --3.52 & Unc \\ 
 & 17554 & 1.06 & 20.10 & 1.08 & -- & -- & -- & 1.86 & --1.55 & Unc \\ 
 & 17591 & 1.59 & 21.17 & 1.43 & -- & -- & -- & 1.86 & --1.12 & Unc \\ 
 & 17642 & 2.48 & 19.94 & 1.41 & -- & -- & -- & 1.86 & --1.61 & Unc \\ 
 & 17683 & 4.41 & 20.77 & 1.04 & -- & -- & -- & 1.86 & --1.28 & Unc \\ 
 & 17703 & 4.83 & 21.37 & 1.4 & -- & -- & -- & 1.86 & --1.04 & Unc \\ 
 & 17806 & 7.53 & 20.95 & 1.24 & -- & -- & -- & 1.86 & --1.21 & Unc \\ 
 & 35115 & 7.54 & 23.2$^{b}$ & -- & -- & -- & -- & 1.86 & --0.56 & Unc \\ 
112 & 15741 & 0.22 & 19.55 & 1.52 & -- & -- & -- & 1.40 & --1.89 & AB \\ 
114 & 35106 & 1.01 & 21.50 & -- & -- & -- & -- & 1.28 & --1.15 & Unc \\ 
115 & 20189 & 0.49 & 19.70 & 1.26 & -- & -- & -- & 1.17 & --1.90 & AB? \\ 
119 & 35107 & 0.61 & 21.50 & -- & -- & -- & -- & 1.17 & --1.19 & Unc \\ 
120 & 20668 & 0.29 & 18.92 & 0.99 & V34 & 0.37274 & EW & 0.96 & --2.31 & NM-AB \\ 
121 & 17091 & 5.45 & 15.585 & 0.999 & -- & -- & -- & 1.03 & --3.61 & Unc \\ 
 & 17190 & 3.63 & 20.52 & 1.23 & -- & -- & -- & 1.03 & --1.63 & Unc \\ 
 & 17201 & 4.08 & 20.40 & 1.08 & -- & -- & -- & 1.03 & --1.68 & Unc \\ 
 & 17218 & 5.76 & 19.65 & 1.41 & -- & -- & -- & 1.03 & --1.86 & Unc \\ 
 & 17219 & 4.99 & 20.63 & 0.87 & -- & -- & -- & 1.03 & --1.59 & Unc \\ 
 & 17221 & 0.53 & 16.730 & 1.106 & -- & -- & -- & 1.03 & --3.15 & Unc \\ 
 & 17261 & 4.73 & 20.46 & 0.76 & -- & -- & -- & 1.03 & --1.66 & Unc \\ 
 & 35108$^{a}$ & 1.56 & 20.43 & -- & -- & -- & -- & 1.03 & --1.67 & Unc \\ 
123 & 15653 & 0.50 & 14.173 & 1.254 & -- & -- & -- & 0.91 & --4.23 & YSS \\ 
124 & 16634 & 0.47 & 19.06 & 1.20 & -- & -- & -- & 0.87 & --2.29 & AB \\ 
125 & 17085 & 0.39 & 17.286 & 1.09 & -- & -- & -- & 0.88 & --3.00 & AB? \\ 
126 & 17601 & 0.11 & 17.331 & 1.141 & V13 & 0.37494 & EW & 0.86 & --2.99 & AB \\ 
127 & 28024 & 0.28 & 18.306 & 1.161 & -- & -- & -- & 1.11 & --2.49 & Unc \\ 
 & 28109 & 3.88 & 21.55 & 1.4 & -- & -- & -- & 1.11 & --1.19 & Unc \\ 
 & 35087 & 3.95 & 22.3 & 1.0 & -- & -- & -- & 1.11 & --0.89 & Unc \\ 
128 & 11434 & 0.57 & 20.26 & 1.42 & -- & -- & -- & 1.02 & --1.74 & AB \\ 
129 & 108-058613 & 0.20 & 13.07 & 0.72 & -- & -- & -- & 0.99 & --4.63 & NM \\ 
130 & 18090 & 0.09 & 15.303 & 0.917 & -- & -- & -- & 0.86 & --3.80 & BSS \\ 
131 & 5597 & 3.46 & 20.07 & 1.32 & -- & -- & -- & 1.14 & --1.77 & Unc \\ 
 & 35091 & 3.69 & 22.2 & 1.0 & -- & -- & -- & 1.14 & --0.91 & Unc \\ 
 & 35092 & 4.96 & 21.68 & 1.4 & -- & -- & -- & 1.14 & --1.13 & Unc \\ 
 & 35109 & 5.71 & 22.1 & -- & -- & -- & -- & 1.14 & --0.96 & Unc \\ 
132 & 23096 & 0.31 & 14.842 & 0.997 & -- & -- & -- & 0.93 & --3.95 & BSS \\ 
133 & 14060 & 0.37 & 19.31 & 1.36 & V10 & 0.3808 & EW & 1.03 & --2.12 & AB \\ 
 & 14035 & 0.86 & 18.858 & 1.21 & -- & -- & -- & 1.03 & --2.30 & AB \\ 
134 & 19489 & 0.76 & 20.54 & 1.61 & -- & -- & -- & 0.84 & --1.71 & AB \\ 
135 & 11985 & 2.17 & 18.112 & 1.188 & -- & -- & -- & 1.55 & --2.42 & Unc \\ 
 & 35110 & 0.32 & 22.3 & -- & -- & -- & -- & 1.55 & --0.73 & Unc \\ \hline
\multicolumn{11}{p {17cm}}{}                                             \\       
\multicolumn{11}{r}{continued on next page}\\
\end{tabular}
\end{center}
\end{table*}

\begin{table*}[h]
\begin{center}
\begin{tabular}{clclccccccc} \hline \hline              
(1) & (2) & (3) & (4) & (5) & (6) &(7) & (8) & (9) &(10) & (11) \\
CX & OID & Dox & $V$ & $B-V$ & Var & P & Variable Type & $L_{X,u}$ &log($F_{X}$/$F_{V}$)$_{u}$ & Class \\
 & & (\arcsec) & & & & (days) & & (10$^{30}$ erg s$^{-1}$) & &  \\ \hline
136 & 15173 & 0.18 & 15.803 & 0.992 & V22 & 1.09430 & EA/EB(BSS) & 0.84 & --3.61 & BSS \\ 
137 & 15517 & 0.45 & 21.65 & 2.1 & -- & -- & -- & 0.71 & --1.35 & NM \\ 
138 & 15696 & 0.50 & 17.810 & 1.15 & V21 & ? & EA? & 0.94 & --2.76 & AB \\ 
139 & 16576 & 0.18 & 19.04 & 1.14 & -- & -- & -- & 0.94 & --2.26 & AB \\ 
142 & 16516 & 0.48 & 17.842 & 1.110 & -- & -- & -- & 0.79 & --2.82 & AB \\ 
143 & 16816 & 1.01 & 20.96 & 1.52 & -- & -- & -- & 0.68 & --1.64 & AB \\ 
144 & 22676 & 0.47 & 16.915 & 0.989 & V24 & 0.35436 & EW & 0.70 & --3.25 & AB \\ 
147 & 18453 & 0.34 & 16.454 & 0.982 & V25 & 0.40091 & EW & 1.43 & --3.12 & AB \\ 
148 & 15852 & 0.05 & 19.81 & 1.59 & -- & -- & -- & 0.65 & --2.12 & AB \\ 
149 & 17650 & 0.63 & 18.962 & 0.72 & V33 & 0.28997 & EW & 0.52 & --2.56 & AB \\ 
150 & 18031$^{c}$ & 2.14 & 16.712 & 1.107 & -- & -- & -- & 0.64 & --3.36 & Unc \\ 
151 & 18292 & 0.45 & 19.16 & 1.07 & -- & -- & -- & 0.51 & --2.49 & CV? \\ \hline
\multicolumn{11}{p {17cm}}{}                                          \\       
\multicolumn{11}{p {17cm}}{Col. 1: X-ray catalogue sequence number } \\
\multicolumn{11}{p {17cm}}{Col. 2: Optical source ID. For five X-ray sources, viz. CX\,7, CX\,30, CX\,40, CX\,100, and CX\,129, the optical source IDs are their UCAC4 catalogue IDs. These stars were saturated in our optical images and their photometry was obtained from the UCAC4 catalogue.}                   \\
\multicolumn{11}{p {17cm}}{Col. 3: Distance between the X-ray source and the optical counterpart in arcsec. } \\
\multicolumn{11}{p {17cm}}{Col. 4: $V$ magnitude, unless specified as $B$ magnitude} \\
\multicolumn{11}{p {17cm}}{Col. 5: $B-V$ colour.}  \\
\multicolumn{11}{p {17cm}}{Col. 6: Short-period binary counterpart ID from \citet{Mazur:1995p95}.}\\
\multicolumn{11}{p {17cm}}{Col. 7: Period (in days) of the short-period binary counterpart.}\\
\multicolumn{11}{p {17cm}}{Col. 8: Variable type, as mentioned in \citet{Mazur:1995p95} }\\
\multicolumn{11}{p {17cm}}{Col. 9: Unabsorbed X-ray luminosity (0.3--7~keV), assuming the source lies at the distance of the cluster, viz. 2.5 kpc.}\\
\multicolumn{11}{p {17cm}}{Col. 10: Unabsorbed X-ray (0.3--7~keV) to optical ($V$ band) flux ratio (2~keV MeKaL model and neutral hydrogen column of 1.9$\times$10$^{21}$cm$^{-2}$)}\\
\multicolumn{11}{p {17cm}}{Col. 11: Object classification : CV? - Candidate cataclysmic variable ; AB(?) - Active binary (candidate) ; SSG - Sub-subgiant ; BSS - Blue straggler star ;  Unc - Uncertain classification; NM - Non-member}\\
\multicolumn{11}{p {17cm}}{Notes: $a$ - photometry of the source may be dubious due to image artefacts (CX81, CX65) or low $S/N$ ratio (CX2, CX24, CX121);  $b$ - the magnitude is a $B$ magnitude; $c$ - the optical counterpart lies just outside the 95$\%$ match radius, but within the 3$\sigma$ match radius; $d$ - value obtained from \citet{Mazur:1995p95}; $e$ - $B$ magnitude obtained from USNO B1.0 catalogue \citep{Monet:2003p984} }         \\
\end{tabular}
\end{center}
\end{table*}

%
% Table 3
%

\begin{table*}[h]
\caption{Properties of close binaries from \citet{Mazur:1995p95} that are matched to a {\em Chandra source}}
\label{tab3}
\begin{center}
\begin{tabular}{cclcccc} \hline \hline              
(1) & (2) & (3) & (4) & (5) & (6) &(7)  \\
CX  & Var & $V$ & $B-V$ & P & Variable Type & log($F_{X}$/$F_{V}$)$_{u}$ \\
 &  & & & (days) &  \\ \hline
18  & V45 & 14.38 &  0.60  & 2.11? & EB(BSS)  & --3.20 \\ 
27  & V42 & 16.37 &  0.98  & 0.7029 & EB   & --2.60 \\ 
68  & V20 & 20.15 &  -- & 0.57712 & EA   & --1.4 \\ 
86  & V12 & 15.81 &  0.69 & 1.4226 & EA(BSS)  & --3.3 \\ 
89  & V11 & 18.9 & 1.18 & 0.5405 & EB/EA & --2.0 \\ 
93  & V30 & 17.02 &  0.86 & 0.35132 & EW  & --2.8 \\ 
97  & V38 & 18.2 &  1.16 & 1.31? & EA  & --2.3 \\ 
120 & V34 & 18.9 & 1.0 & 0.37274 & EW  & --2.3 \\ 
126 & V13 & 17.29 &  1.00 & 0.37494 & EW & --3.0 \\ 
133 & V10 & 19.3 &  -- & 0.3808 & EW & --2.1 \\ 
136 & V22 & 15.9 &  0.73 & 1.09430 & EA/EB(BSS) & --3.6  \\ 
138 & V21 & 17.74 &  1.20 & ? & EA? & --2.8 \\ 
144 & V24 & 16.84 &  1.1 & 0.35436 & EW & --3.3  \\ 
147 & V25 & 16.38 &  0.88 & 0.40091 & EW & --3.1 \\ 
149 & V33 & 18.24 &  1.09 & 0.28997 & EW & --2.8\\ \hline
\multicolumn{7}{p {10cm}}{Col. 1: X-ray catalogue sequence number. } \\
\multicolumn{7}{p {10cm}}{Col. 2: Short-period binary counterpart ID from \citet{Mazur:1995p95}.}                   \\
\multicolumn{7}{p {10cm}}{Col. 3 and 4: $V$ magnitude and $B-V$ colour from \citet{Mazur:1995p95}.} \\
\multicolumn{7}{p {10cm}}{Col. 5: Period (in d) of the short-period
  binary counterpart from \citet{Mazur:1995p95}.}\\
\multicolumn{7}{p {10cm}}{
  Col. 6: Variable type, as mentioned in \citet{Mazur:1995p95}: EA =
  eclipsing binary of the Algol type, EB = $\beta$
  Lyrae type variables with unequal minima and maxima in the light
  curve, and EW is a contact binary of the W\,UMa type.}\\
\multicolumn{7}{p {10cm}}{Col. 7: Unabsorbed X-ray (0.3--7~keV) to
  optical ($V$ band) flux ratio (2~keV MeKaL model and neutral
  hydrogen column of $1.9\times10^{21}$ cm$^{-2}$).}\\

\end{tabular}
\end{center}
\label{tab_mazur}
\end{table*}

%
% Table 4
%

 \begin{table*}[h]
 \caption{Comparison among old open clusters of X-ray sources with $L_X\geq10^{30}$ erg s$^{-1}$ (0.3--7~keV) inside $r_h$}
 \label{tab5}
\begin{center}
 \begin{tabular}{lccccccc} \hline \hline              
Cluster        &  Age       & Mass               & $N_X$ &  $N_{X,CV}$& $N_{X,SSG}$& $N_{X,AB}$  & log(2$L_{30}$/Mass) \\ 
               & (Gyr)      & ($M_{\odot}$)       &        &  & & &\\ \hline
 NGC\,6819$^2$ & 2 -- 2.4     & 2600               & 3 -- 8 & $\lesssim1$ & $\lesssim1$ & $\lesssim4$ & 28.8--29.3 \\ 
 M\,67$^1$     & 4          & 2100$^{+610}_{-550}$ & 12      & 0 & 1 & 7 -- 8 & 28.6  \\
 NGC\,6791$^3$ & 8          & 5000--7000         & 15 -- 19 & 3 -- 4 & 3 & 7 -- 11 & 28.6 -- 28.8\\
 Cr\,261       & 7          & 5800--7200         & $\lesssim$26$\pm$8 & $\lesssim4$ & $\lesssim2$ & $2 - 23$ & $\lesssim$28.6 -- 28.7 \\ \hline
 
\multicolumn{8}{l}{}                                             \\       
\multicolumn{8}{p {11cm}}{Col.\ (1): Cluster name listed in order of increasing age}         \\
\multicolumn{8}{p {11cm}}{Col.\ (2): Cluster age in Gyr}         \\
\multicolumn{8}{p {11cm}}{Col.\ (3): Cluster mass in $M_{\odot}$. The
  estimate for Cr\,261 is based on the integrated $V$ magnitude of the
  cluster (Sect.~\ref{sec_clusterprop}); for the other clusters, see
  the references quoted below.}\\
\multicolumn{8}{p {11cm}}{Col.\ (4): Number of X-ray sources inside
  $r_h$ with $L_X \geq 1\times10^{30}$ erg s$^{-1}$} \\
\multicolumn{8}{p {11cm}}{Col.\ (5): Number of candidate CVs inside
$r_h$ with $L_X \geq 1\times10^{30}$ erg s$^{-1}$. M\,67 does host the CV EU\,Cnc inside $r_h$, but it is fainter
than the luminosity cutoff.} \\
\multicolumn{8}{p {11cm}}{Col.\ (6): Number of candidate SSGs inside $r_h$ with
  $L_X \geq 1\times10^{30}$ erg s$^{-1}$} \\
\multicolumn{8}{p {11cm}}{Col.\ (7): Number of (candidate) ABs inside
  $r_h$ with $L_X \geq 1\times10^{30}$ erg s$^{-1}$. The lower limit to the number of ABs in Cr\,261 is set by CX\,93\,V30 and CX\,147/V25, two W\,UMa's at a distance that is consistent with that of the cluster.} \\

\multicolumn{8}{p {11cm}}{Col.\ (8): Ratio of the total X-ray
  luminosity of sources inside $r_h$ brighter than $1\times10^{30}$ erg
  s$^{-1}$ ($L_{30}$), and cluster mass. The multiplicative factor 2
  is included to scale the mass estimate to the half-mass radius. The
  value for M\,67 has been updated with respect to the
  \citet{vandenBerg:2013p442} value, to account for an updated mass
  estimate \citep{gellea15}. For NGC\,6819, new membership information
  from \citet{platea11} has been included.} \\

\multicolumn{8}{p {11cm}}{References --
  $^1$\citet{vandenBerg:2004p1040}, \citet{gellea15},
  $^2$\citet{Gosnell:2012p685}, \citet{platea13}
  $^3$\citet{vandenBerg:2013p442}, \citet{platea11}}

      \end{tabular}
  \end{center}
\end{table*}

\end{document}